\documentclass[a4paper, 10pt]{article}

\usepackage{geometry}
\usepackage[utf8]{inputenc}
\usepackage[english]{babel}
\usepackage{setspace}
\usepackage{lipsum}
\usepackage{xcolor}
\usepackage{hyperref}
\hypersetup{
    colorlinks=false
    }
        
\usepackage[]{natbib}
\usepackage{amsfonts}
\usepackage{amsmath}
\usepackage{caption}
\usepackage{authblk}
\usepackage{graphicx}
\graphicspath{{figures/}}
\usepackage[caption=false]{subfig}
\usepackage{appendix}
\usepackage{comment}
\usepackage{array}

\providecommand{\keywords}[1]
{
  \small	
  \textbf{Keywords:} #1
}

\begin{document}

\title{Non-linear stability analysis of slip in a single-degree-of-freedom elastic system with frictional evolution laws spanning aging to slip}

\author[1]{Federico Ciardo}
\author[1]{Robert C. Viesca}

\affil[1]{Department of Civil and Environmental Engineering, Tufts University, Medford, MA, USA.}

\date{}

\noindent
\begin{center} 
\large{Non-peer reviewed preprint submitted for consideration}
\end{center}

\begingroup
\let\newpage\relax
\maketitle
\endgroup


\begin{abstract}
We present a non-linear stability analysis of quasi-static slip in a spring-block model. The sliding interface is governed by rate- and state-dependent friction, with an intermediate state evolution law that spans between aging and slip laws using a dimensionless parameter $\epsilon$. Our results extend and generalize previous findings of \citet{GuRi84} and \citet{RaRi99} that considered slip and aging laws, respectively. We examine the robustness of these prior results to changes in the evolution law, including the finding of unconditional stability of the aging law for spring stiffnesses above a critical value. Our analysis provides analytical trajectories of slip motion in a phase plane as function of dimensionless governing parameters. We investigate two scenarios: a spring-block model with stationary and non-stationary point loading rate.
When the loading point is stationary, we find that deviations from the aging law lead to only conditional stability of the slider for spring stiffnesses above a critical value: finite perturbations can trigger instability, consistent with prior results for the slip law. We quantify these critical perturbations as a function of the governing parameters. We find that, for a given supercritical stiffness, the size of the perturbation required to induce instability grows as the state evolution law approaches the aging law. In contrast, when the point loading rate is stationary, our results suggest that there exists a maximum critical stiffness above which an instability can never develop, for any perturbation size. This critical stiffness is $\epsilon$-dependent and vanishes as the slip law is approached $\left(\epsilon \to 0\right)$: conditional stability is then expected in the slip law limit. For a limited extension of the results found for a spring-slider to continuum faults, we derive relations for an effective spring stiffness as a function of the elastic moduli and a characteristic fault dimension or a characteristic perturbation wavelength. 

\end{abstract}


\section*{Keypoints}
\begin{itemize}
\item We examine the motion and stability of a single-degree-of-freedom elastic system with a rate- and state-dependent friction.
\item We allow for the state evolution law to differ from two previously considered laws, allowing to test the robustness of prior results. 
\item We find that unconditional stability of the aging law for large stiffnesses is specific only to that law.
\item Extending our results to a continuous fault, our results suggest that nucleation size for a dynamic instability can generally be expected to be dependent on the strength of the perturbation.
\end{itemize}

\section{Introduction}
\label{sec:introduction}

\subsection{Motivation}

Stability of slip on frictional interfaces in response to external perturbations continues to pose a significant research challenge. The conditions dictating the transition between stable and unstable frictional sliding remain elusive for a range of realistic descriptions of frictional strength.

Given the laboratory evidence that shows the dependence of friction on both sliding velocity and time of nominally stationary contact \citep[e.g.,][]{Diet79, Ruin83}, rate- and state-dependent frictional models have emerged as a preferred framework for theoretical investigations of slip stability. However, these investigations have been mostly limited to specific state evolutions laws. Here, we perform a non-linear stability analysis of quasi-static slip using a simplified spring-block elastic model where frictional strength is parametrized to encompass a range of possible frictional evolution laws, and quantify the conditions that trigger a slip instability, addressing questions regarding the generality of prior results.

The non-linear coupling between friction and elasticity along a slipping interface gives rise to elasto-frictional length-scales that scale critical dimensions, determining slip dynamics. Theoretical models of a fault in an elastic continuum with rate-and-state friction and the so-called aging-law state evolution demonstrated the emergence of a quasi-static nucleation phase in which slip unstably accelerates within a compact nucleation distance along the fault, after which inertia inhibits further acceleration in this region, leading to outward dynamic rupture \citep{Diet92, RuAm05, Vies16a, Vies16b}. Several laboratory experiments of frictional instabilities tend to support such a nucleation process e.g. \citep{OkuDiet84,OhnaShe99,NiTa10,LaScu13,PaLa17,McLaskey19,GviFin21,MaSch2023}.
However, other theoretical models based on a different state evolution law, the slip law (which is known to give a different frictional behavior in response to changes in slip rates that seems to best reproduce laboratory observations \citep{BhaRu22}) revealed that this critical dimension for dynamic slip nucleation is not uniquely defined. Also, using a fault in an elastic continuum model, \citet{AmRu08} demonstrated that under the slip law, the nucleation zone is comparatively smaller, continually shrinking as slip unstably accelerates. More recently, \citet{Viesca23} further showed that this nucleation zone can vanish for slip-law frictional response, implying that a point-wise divergence of slip rate can theoretically take place. These results are consistent with previous findings obtained by \citet{GuRi84} using a spring-block model with rate-and-state friction and the same state evolution law.
Their research highlighted that a slip instability can be triggered at any value of spring stiffness (which inversely correlates with the dimension of a finite slipping fault, to be demonstrated in sub-section \ref{subsec: analogy spring-block continuum}), provided that a sufficiently large perturbation is applied to the elastic system. 

Perturbations in loading rate can be localized in space, hence in the form of rapid fluid injection in a fault or in the form of creep converging on a stuck brittle asperity, or can be more distributed but instantaneous in time, for instance as a result of elastic stress transfer onto one fault due to the sudden slip of another fault nearby. The question whether the nucleation process of a dynamic instability is sensitive to the applied loading rate has been the subject of many laboratory studies. Injection-induced rupture initiation and propagation on laboratory faults have consistently revealed that fast injection rates lead to dynamic rupture nucleation over spatial scales much smaller than the quasi-static theoretical estimates of nucleation sizes \citep{GoRu21, JiWa22}. Similar results have also been obtained on large-scale laboratory faults when subjected to sufficiently fast background loading rates \citep{XuFu18,GueNie19}, or when ``kicked" by a rapid increase of loading rate after a hold period (as opposed to the spontaneous nucleation upon an increasing of loading rate with no healing period) \citep{McLaYama17,McLaskey19}. This sensitivity of nucleation size on loading rate is consistent with theoretical and numerical models of continuum faults obeying the slip law (e.g, see \citep{KaNie16, Ga21, Cattania23}). 

Among the questions we look to address is how general is such loading-rate sensitivity? Do spring-block models indicate that a critical nucleation length is uniquely defined when frictional state deviates, even slightly, from the aging law?
Can we quantify the critical perturbation in loading that can trigger a dynamic instability for different frictional strength evolution laws? We use the simplest model to study slip motion on frictional interfaces of compliant bodies, i.e. the single-degree-of-freedom spring-slider elastic system depicted in Figure \ref{fig: spring-slider}. We investigate slip stability in response to sudden, finite-strength perturbations in either the slider's displacement or point loading rate. A family of state evolution laws is considered using a so-called intermediate law that ranges between the slip law to the aging law. We draw analogies between the spring-block model and faults in continua and discuss the extent to which results may be extrapolated from the former to the latter. Given these analogies of a discrete elastic model and a continuum fault model, here presented in details for some particular configurations, the findings of this study have application to continuous frictional interfaces.

\subsection{Rate-and-state friction}
\label{subsec: RateAndState}
Rate-and-state friction, originally proposed by \citet{Diet78, Diet79, Ruin83}, has been widely used to study stick-slip instabilities resulting from elastic bodies in frictional contact. Driven by laboratory friction experiments of stick-slip motion that were thought to be analogous to earthquake ruptures in geological faults, rate-and-state friction law has been developed based on some experimental observations: i) the evolution of the friction coefficient from a ``static" value when no slip occurs, down to a ``dynamic" value when slip (or slip rate) is non-zero \citep{Byerlee1970}, ii) the dependence of ``static" friction stress on time of nominally stationary contact, iii) the dependence of friction resistance on instantaneous slip rate, iv) the dependence of friction on recent slip and finally v) the steady-state feature of friction stress associated with any uniform slip \citep{Ruin83}.
The phenomenological description of all these phenomena has led to the so-called rate-and-state friction law, in which the friction coefficient $f$ depends on both sliding velocity $V$ and state variable $\theta$. Mathematically, this law reads
\begin{equation}
f\left( V, \theta \right) = f_* + a \ln \frac{V}{V_*} + b \ln \frac{V_* \theta}{D_c},
\label{eq: rate and state friction}
\end{equation}
where $a$ and $b$ are two phenomenological constants of often inferred to be order 0.01, $D_c$ is a characteristic slip distance, and finally $f_*$ and $V_*$ are the value of friction coefficient and slip rate at reference steady-state for which $\theta = D_c/V_*$. Here $f$ is a Mohr-Coulomb friction coefficient relating the interfacial shear stress $\tau$ to the interfacial normal stress $\sigma$ as $\tau = f \sigma$.
The phenomenological evolution equation for $\theta$ can be generally casted in the following form \citep{Ruin83}
\begin{equation}
\frac{\partial \theta}{\partial t}  = G \left( V, \theta\right),
\end{equation}
where the function $G$ is assumed to be such that the state variable evolves towards a steady-state value $\theta_{ss} \left( V \right)$ after sufficient time or  displacement, satisfying $G=0$ at constant slip rate $V$. By considering the state variable $\theta$ as the weighted average of some function of the recent history of slip rate $g\left( V/V_c\right)$ (with $V_c$ being a given dimensionless constant), \citet{Ruin83} proposed different state evolution laws in his seminal work. If the function being averaged is the slowness $g\left( V/V_c\right) = V_c/V$, the state variable varies in time even at null slip rates and its corresponding time evolution follows the so-called ``aging" law \citep{Diet78, Diet79, Ruin83}:
\begin{equation}
\frac{\partial \theta}{\partial t} = 1 - \frac{V \theta}{D_c} \qquad \text{(Dieterich-Ruina aging law)}
\label{eq: aging law}
\end{equation}
Alternatively, if $g\left( V/V_c\right) = \ln (V/V_c)$, the state variable evolves solely with slip and its time evolution follows the so-called ``slip" law:
\begin{equation}
\frac{\partial \theta}{\partial t} =  - \frac{V \theta}{D_c} \ln \frac{V \theta}{D_c}  \qquad \text{(Ruina-Dieterich slip law)}
\label{eq: slip law}
\end{equation}

Over the past few decades, the aging law (\ref{eq: aging law}) has been the law of choice in many frictional-related theoretical and numerical studies, partly because of the inability of slip law (\ref{eq: slip law}) to account for the time-dependent healing of frictional surface during stationary $(V=0)$ contact. More recent analysis of laboratory data, however, suggests the friction coefficient is more sensitive to slip than elapsed time, even at nearly null slip velocities, implying that the description that best reproduces observations is one in which slip is required for frictional strength to evolve \citep{BhaRu17, BhaRu22}. 

Both of these laws are asymptotically identical (to leading order) in the vicinity of steady-state conditions, during which $\partial \theta / \partial t =0$ and hence $\dfrac{V \theta}{D_c}=1$.
Away from steady-state, these two laws will differ, and, for example, are known to lead to different frictional behavior in response to sudden increases or decreases of slip velocity. Under the slip law a symmetric response to velocity changes occurs, while the corresponding response under the aging law is asymmetric (see for instance Figure \ref{fig: velocity stepping experiments}).
The transition between these two different responses can be captured using an intermediate state evolution law that bridges the slip and aging laws. For example, if $\theta$ is considered as an average of a power of $V$, i.e. $g\left( V/V_c\right) = \left(V/V_c \right)^{-\epsilon}$ with $\epsilon$ being a dimensionless parameter, the time evolution of the state variable follows
\begin{equation}
\frac{\partial \theta}{\partial t} = \frac{1}{\epsilon} \left[ \left( \frac{V \theta}{D_c}\right)^{-\epsilon} - 1\right] \frac{V \theta}{D_c}, \qquad \text{(Intermediate state evolution law)}
\label{eq: intermediate state evolution law}
\end{equation}
with the end-member aging (\ref{eq: aging law}) and slip (\ref{eq: slip law}) laws retrieved respectively with $\epsilon = 1$ and $\epsilon \to 0$. This intermediate state evolution law was also proposed by \citet{Ruin83}, but has received little attention until recently \citep{Viesca23, NoCha23}.

\subsection{Prior linear and non-linear stability analyses}
\label{subsec: linear and non-linear stability analysis}
\begin{figure}[t!]
\centering
\includegraphics[width=.8\textwidth]{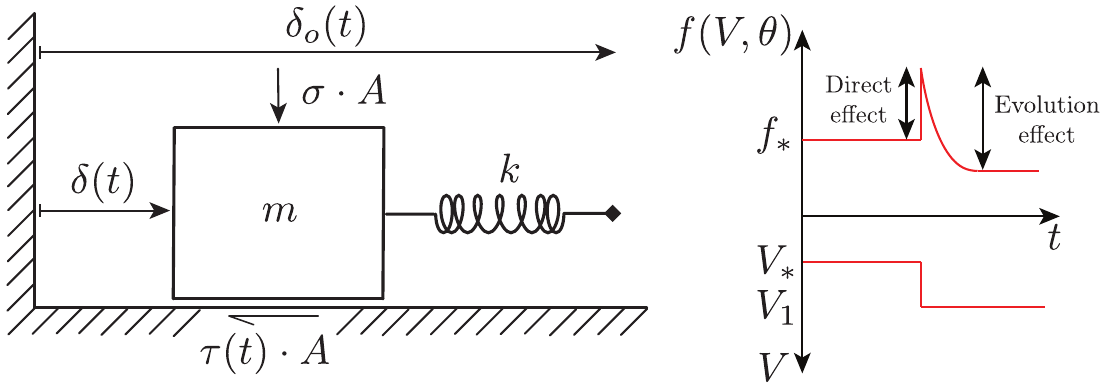}
\caption{Left: Single-degree-of-freedom spring-slider elastic system. The slider of mass $m$ moves along the frictional interface governed by rate- and state-dependent friction. It slides a distance $\delta$, pulled at one extremity of the spring with characteristic stiffness $k$ at a displacement rate $V_o = d \delta_o/dt$. Inertia is neglected, i.e. the product $m \cdot \dot{V}$ is negligible. Right: Frictional response to step change in sliding velocity. The friction coefficient experiences an instantaneous increase from reference value $f_*$ upon velocity jump (direct effect), before evolving to a new steady-state value over a characteristic slip distance $D_c$ (evolution effect). In this contribution, the new value of friction coefficient is always lower than the reference value, i.e. a steady-state rate-weakening condition is considered.}
\label{fig: spring-slider}
\end{figure}

Using a single-degree-of-freedom spring-slider elastic system, such as the one depicted in Figure \ref{fig: spring-slider}-left, and the previously described rate-and-state description of frictional sliding, many efforts have been devoted to study frictional slip in response to external perturbations and the conditions that govern stable and unstable frictional sliding \citep[e.g][]{RiTse86,GuWo91,HeSa09}. 

\citet{Ruin83} used linear stability analysis to show that the response to small perturbations about steady-state sliding is unstable provided that the spring stiffness $k$ is lower than a critical value defined as
\begin{equation}
k_{crit} = \frac{\sigma \left(b-a\right)}{D_c}
\label{eq: critical stiffness}
\end{equation}
This condition holds with both slip and aging law (as well as with the intermediate law), owing to their common linearization about steady state.
When the frictional non-linearity becomes important following the growth of small perturbations, the form of the evolution law impacts the subsequent development of the instability \citep{GuRi84, RaRi99, AmRu08, Viesca23}.
Using a spring-slider model with the slip law (\ref{eq: slip law}), \citet{GuRi84} showed that an no inertia instability can nucleate with any value of spring stiffness $k$, provided that sufficiently large perturbations are applied to the system (with larger finite-strength perturbations needed for values of $k$ increasing above $k_{crit}$). 
This lies in contrast to the results of a subsequent non-linear stability analysis using the aging law (\ref{eq: aging law}), in which unconditional stability was found for $k>k_{crit}$ \citep{RaRi99}.

Given the simplifications inherent in assuming a single state variable description, in addition to the assumption that a particular form of that description is correct, one is left to wonder whether either of these results (unconditional or conditional stability) are robust to perturbations of the friction law itself. To test this, we extend the non-linear stability analysis done for the aging (\ref{eq: aging law}) and slip (\ref{eq: slip law}) laws to the intervening family of laws captured by the intermediate law (\ref{eq: intermediate state evolution law}).
Although here we use a simple single-degree-of-freedom frictional model, our results have extensions to a frictional interface of continuous, deformable bodies, provided that mappings from a discrete spring-slider to a continuum interface model are known. In the following sub-section we provide a concise description of these mappings.

\subsection{Mapping a spring-block model to a continuous frictional interface}
\label{subsec: analogy spring-block continuum}

\begin{figure}[t!]
\centering
\includegraphics[width=.75\textwidth]{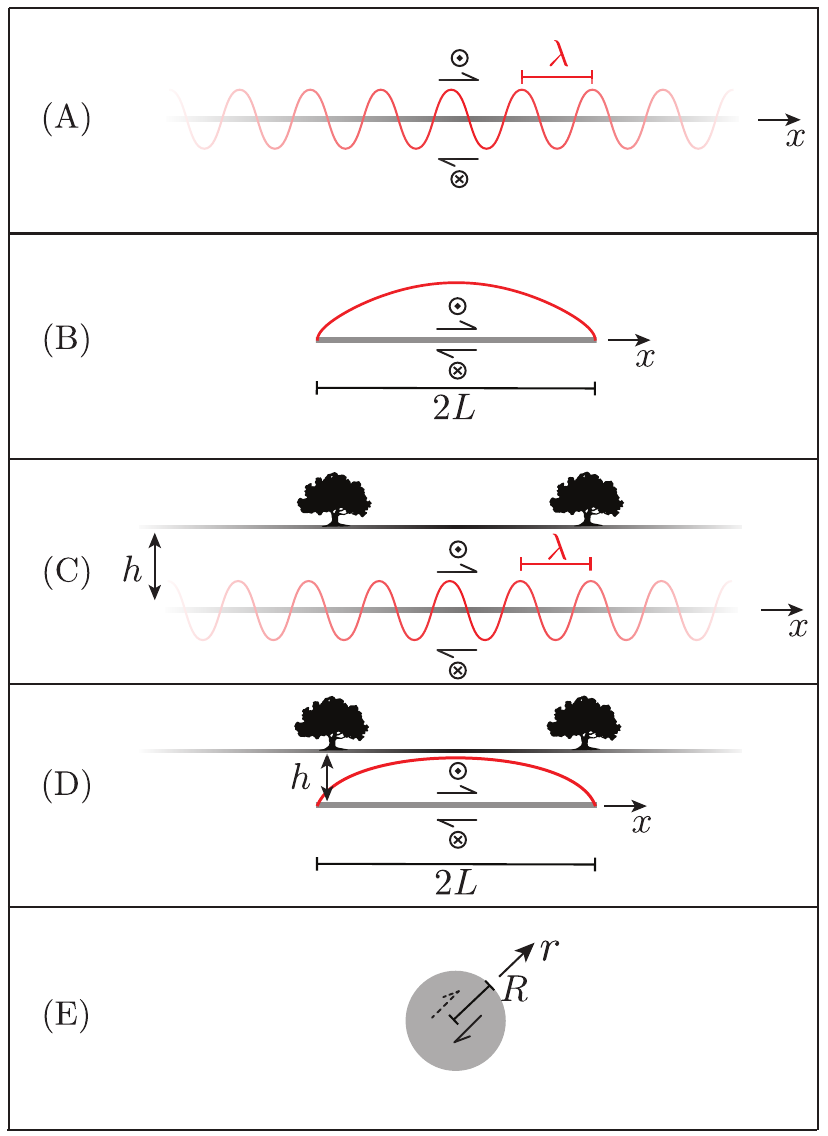}
\caption{Different configurations and geometries for a deformable frictional interface (i.e. frictional fault). Cases (A) and (C) represent respectively the scenario of an infinite planar fault embedded within an unbounded linear-elastic medium and within a half-space, with its distance to the free-surface denoted by $h$. Cases (B) and (D) are respectively equivalent to cases (A) and (C), with the only difference that the frictional fault has a finite length $2L$. Finally, case (E) denotes a circular fault embedded within a three-dimensional unbounded medium.}
\label{fig: continuum models}
\end{figure}

The spring-block model can be considered as an idealization of a continuous interface model, approximating the latter model with its infinite degrees of freedom to one with a single degree of freedom. 

On a continuous fault, we may decompose the shear stress as 
\begin{equation}
\tau \left(x,t\right) = \tau_{el} \left( x,t\right) + \tau_{ext} \left( x,t\right),
\end{equation}
where, for quasi-static motion, $\tau_{el}\left( x,t\right)$ is the instantaneous change in shear stress along the fault owed to a continuous distribution of slip $\delta_c \left( x,t\right)$ and is determined by a linear operator acting on the spatial component of $\delta$ 
\begin{equation}
\tau_{el} = \mathcal{L} \left[ \delta \left(x,t \right)\right],
\end{equation}
and $\tau_{ext} \left( x,t\right)$ is the shear stress on the surface in the absence of slip that may be considered as an external forcing term. To find an equivalent stiffness for the continuum fault, we consider an idealized scenario where slip evolves as 
\begin{equation}
\delta_c \left(x,t\right) = D\left(t\right) \cdot \delta\left( x\right)
\end{equation}
and subsequently look to find a stiffness $k$ satisfying $\tau_{el} = k \cdot \delta$, in analogy to a spring force $F_{el} = k \cdot \delta$. Given that the operator $\mathcal{L}$ has the property that $\mathcal{L}\left[ g(t) \delta(x)\right] = g(t) \mathcal{L}\left[\delta(x)\right]$, finding this stiffness $k$ reduces to the solution of the eigenvalue problem 
\begin{equation}
k \cdot \delta\left( x \right) = \mathcal{L} \left[ \delta(x)\right]
\label{eq: eigenequation}
\end{equation}
with values of $k$ representing eigenstiffnesses with corresponding mode shapes $\delta(x)$. The operator $\mathcal{L}$ depends of the mode of rupture. For some particular configurations the eigenvalue problem (\ref{eq: eigenequation}) can be solved analytically, but for others it can only be solved numerically or by asymptotic approximation. 
In Table \ref{tab: table} we include a few cases, for each of which the corresponding expressions of the elastic operator $\mathcal{L}(\delta)$ and the normalized stiffness $k/k_*$ are reported. 
Figure \ref{fig: continuum models} shows the corresponding schematics of the continuum configurations, including the problem domain (bounded or unbounded), the mode of slip (in-plane, anti-plane, or mixed-mode), and presence or not of a free surface.
For details on how the values of $k/k_*$ are obtained, we refer the reader to Appendix \ref{app: appendix 2}.

\begin{table}[t!]
\centering
    \begin{tabular}{| m{0.18\linewidth} | m{0.44\linewidth}| m{0.14\linewidth}| m{0.13\linewidth}|}
    \hline
      \centering \textbf{Continuous configuration}  &\centering  \textbf{Elastic operator $\mathcal{L}(\delta)$} & \centering \textbf{Characteristic stiffness $k_*$} & \centering\arraybackslash \textbf{$k/k_*$} \\ \hline
      \centering (A) Infinite interface between two unbounded elastic half-spaces &  \[ \mathcal{L}\left( \delta\right) = \dfrac{\mu^{\prime}}{\pi} \int_{-\infty}^{\infty} \dfrac{\partial \delta/\partial s}{x-s} \,\text{ds} \] 
      & \centering $\dfrac{\mu^{\prime}}{\lambda}$ &   \centering\arraybackslash $2 \pi$ \\ \hline
      \centering (B) Finite-length interface in an unbounded elastic medium &\centering  \[ \mathcal{L}\left( \delta\right) = \dfrac{\mu^{\prime}}{\pi} \int_{-L}^{L} \dfrac{\partial \delta/\partial s}{x-s} \,\text{ds} \] &  \centering $\dfrac{\mu^{\prime}}{L}$ &  \centering\arraybackslash $1.157773882\dots$ (Smallest)  \\ \hline
      \centering (C) Infinite interface near and parallel to a free-surface of an unbounded half-space & \centering \[ \mathcal{L}\left( \delta\right) =  -E^{\prime} h \dfrac{\partial ^{2} \delta}{\partial x^{2}}, \quad -\infty < x < \infty, \quad h\ll \lambda \]  & \centering $\dfrac{E^{\prime} h}{\lambda^2}$ &  \centering\arraybackslash $4 \pi^2$  \\ \hline
      \centering (D) Finite-length interface near and parallel to a free-surface of an unbounded half-space & \centering \[ \mathcal{L}\left( \delta\right) =  -E^{\prime} h \dfrac{\partial ^{2} \delta}{\partial x^{2}}, \quad -L \leq x \leq L, \quad h \ll L \]& \centering  $\dfrac{E^{\prime} h}{L^2}$ &  \centering\arraybackslash $\pi^2/4$ (Smallest) \\ \hline
      \centering (E) Circular interface in an unbounded elastic medium ($\nu = 0$) & \centering \[ \mathcal{L}\left( \delta\right) = \dfrac{\mu}{2 \pi} \int_{0}^{R} \dfrac{\partial \delta}{\partial s} \left( \dfrac{E\left[ f(r/s)\right]}{r-s} - \dfrac{F\left[ f(r/s)\right]}{s+r}\right)  \text{ds} \] & \centering $\dfrac{\mu}{R}$ &  \centering\arraybackslash $ 1.003059516\dots$ (Smallest) \\
      \hline
    \end{tabular}
  \caption{Normalized stiffness $k/k_*$ for different continuum frictional interface models (see Appendix \ref{app: appendix 2} for more details). For cases (B), (D) and (E), only the smallest eigenstiffness is reported. For each problem configuration and geometry, the corresponding elastic operator $\mathcal{L}(\delta)$ is reported. $\mu^{\prime}$ is an effective shear modulus that depends on slip-mode: $\mu^{\prime} =  \mu/2(1-\nu)$ in Mode II and $\mu^{\prime} = \mu/2$ in mode III, where $\mu$ is the shear modulus and $\nu$ is the Poisson's ratio. Similarly, $E^{\prime}$ is an effective slip-mode-dependent modulus that is given by $E^{\prime} = 2 \mu/(1-\nu)$ in mode II and $E^{\prime} = \mu$ in mode III. Finally, in cases (C) and (D), $h$ represents the depth of the interface from free-surface (see also Figure \ref{fig: continuum models}). The condition ``near" the free-surface is guaranteed if $h$ is much smaller than the applied perturbation wavelength $\lambda$ (for case (C)) or the fault half-length $L$ (for case (D)).}
  \label{tab: table}
\end{table}

In Table \ref{tab: table}, we note that on unbounded frictional interfaces the stiffness is inversely related to the applied perturbation wavelength $\lambda$. In this case, the eigenvalues are continuous and thus the stiffness varies continuously with $\lambda$. For example, for in-plane or anti-plane slip at interface of two unbounded half-spaces, the stiffness reads 
\begin{equation}
k = 2\pi \cdot \dfrac{\mu^{\prime}}{\lambda},
\end{equation}
where $\mu^{\prime}$ is an effective slip-mode-dependent shear modulus: $\mu^\prime = \mu/2(1-\nu)$ in mode II and $\mu^\prime/2$ in mode III, with $\nu$ being the Poisson's ratio. 
In contrast, for finite-sized interfaces, the stiffness varies inversely with the surface length (here half-length $L$ or radius $R$). Moreover, the finite problem domain leads to discrete, but countably infinite eigenvalues. For example, for in-plane or anti-plane slip along an interface of length $2L$ embedded within a continuum, we may write the discrete set of eigenstiffness as 
\begin{equation}
k_i = \eta_i \cdot \frac{\mu^{\prime}}{L} \qquad i = 0,1,2,\dots
\end{equation}
where $\eta_0  = 1.157773882\dots$ is the leading eigenvalue. As we are concerned with the lowest possible stiffness of a system with given $\mu^{\prime}$ and $L$ (or $R$), to compare against the critical stiffness $k_{crit}$, the leading eigenvalue is the salient one. 
The presence of a critical stiffness in a spring-block elastic system, for which slip is always unstable when $k<k_{crit}$ regardless of the strength of the perturbation applied \citep{Ruin83,GuRi84,RaRi99}, therefore suggests that in a continuum model there exists a critical wavelength, or a critical fault dimension, above which small perturbations to uniform steady-state sliding grow in time. These latter can be determined by equating the expression for $k_{crit} $(\ref{eq: critical stiffness}) with the values of $k$ reported in Table \ref{tab: table}, obtaining for each scenario a critical dimension $\left( \lambda_c, L_c,\, \text{or}\, R_c\right)$
\begin{align}
\begin{split}
\lambda_c &= 2 \pi \cdot \frac{L_b}{1-a/b} \quad \text{Scenario (A)} \\
L_c &\simeq \left( 1.157773882\dots\right) \cdot \frac{L_b}{1-a/b} \quad \text{Scenario (B)}\\
\lambda_c &= 2 \pi \cdot L_{bh} \sqrt{\dfrac{1}{1-a/b}} \quad \text{Scenario (C)}\\
L_c &= \dfrac{\pi}{2} \cdot L_{bh}  \sqrt{\frac{1}{1-a/b}} \quad \text{Scenario (D)}\\
R_c &\simeq \left( 1.003059516\dots \right) \cdot \frac{L_b}{1-a/b} \quad \text{Scenario (E)}
\label{eq: critical dimensions}
\end{split}
\end{align}
where $L_b = \dfrac{\mu^{\prime} D_c}{\sigma b}$ and $L_{bh} = \sqrt{\dfrac{E^{\prime} h D_c}{\sigma b}}$ are two elasto-frictional length scales (with $\mu^{\prime} = \mu$ in scenario (E)). 

As a final note, we point out the caveat that decomposing the response of a continuum model into independent eigenmodes is only asymptotically correct in the linear regime. Once frictional non-linearities are no longer negligible, coupling of modes occurs, and a distinct analysis of instability in continuous faults is required \citep[e.g][]{RuAm05, Vies16a,Vies16b,Viesca23}.

\section{Equations of motion for spring-slider elastic system}
\label{sec: equations of motion}

We consider the single-degree-of-freedom spring-slider elastic system depicted in Figure \ref{fig: spring-slider} that undergoes frictional slip.
The slider is in contact with an elastic body and moves at a velocity $V  = \dot{\delta} = d \delta/d t$, 
pulled by a point loading rate $V_o = d \delta_o/d t$ imposed at one extremity of the spring. The spring is characterized by a stiffness $k$ (per unit area of sliding contact) whose magnitude modulates the elastic force transmitted to the slider during slip motion. Normal stress $\sigma$ and frictional law constitutive parameters $a$ and $b$ are assumed to remain constant on the contact interface. Motion is assumed to be quasi-static and the block's inertia is neglected: i.e. the product $m \cdot \dot{V}$ is neglected, with $m$ being the block mass.

The equation of motion of the single-degree-of-freedom elastic system is given by the following equilibrium equation
\begin{equation}
F_{el} + F_{f} - F_{ext} = 0,
\label{eq: equation of motion1}
\end{equation}
where $F_{el} = k \cdot \delta$ is taken to be the elastic force due to slip defined as such the single-degree of freedom counterpart to $\tau_{el}$ in the continuum model, $F_{ext} = k \cdot \delta_o$ is the external force proportional to the imposed displacement $\delta_o$ and likewise the single-degree of freedom counterpart to $\tau_{ext}$, and $F_{f} = \tau \cdot A$ is the frictional force along the interface, which can be expressed, assuming a unit base area, as the product of rate- and state-dependent friction coefficient (\ref{eq: rate and state friction}) and applied normal stress $\sigma$:
\begin{equation}
F_{f} = \tau = f\left( V, \theta \right) \times \sigma
\label{eq: MC criterion}
\end{equation}
In rate form, Equation (\ref{eq: equation of motion1}) reads
\begin{equation}
\frac{d \tau}{d t} = k \left( V_o - V\right)
\label{eq: equation of motion2}
\end{equation}

\begin{figure}[t!]
\centering
\includegraphics[width=.9\textwidth]{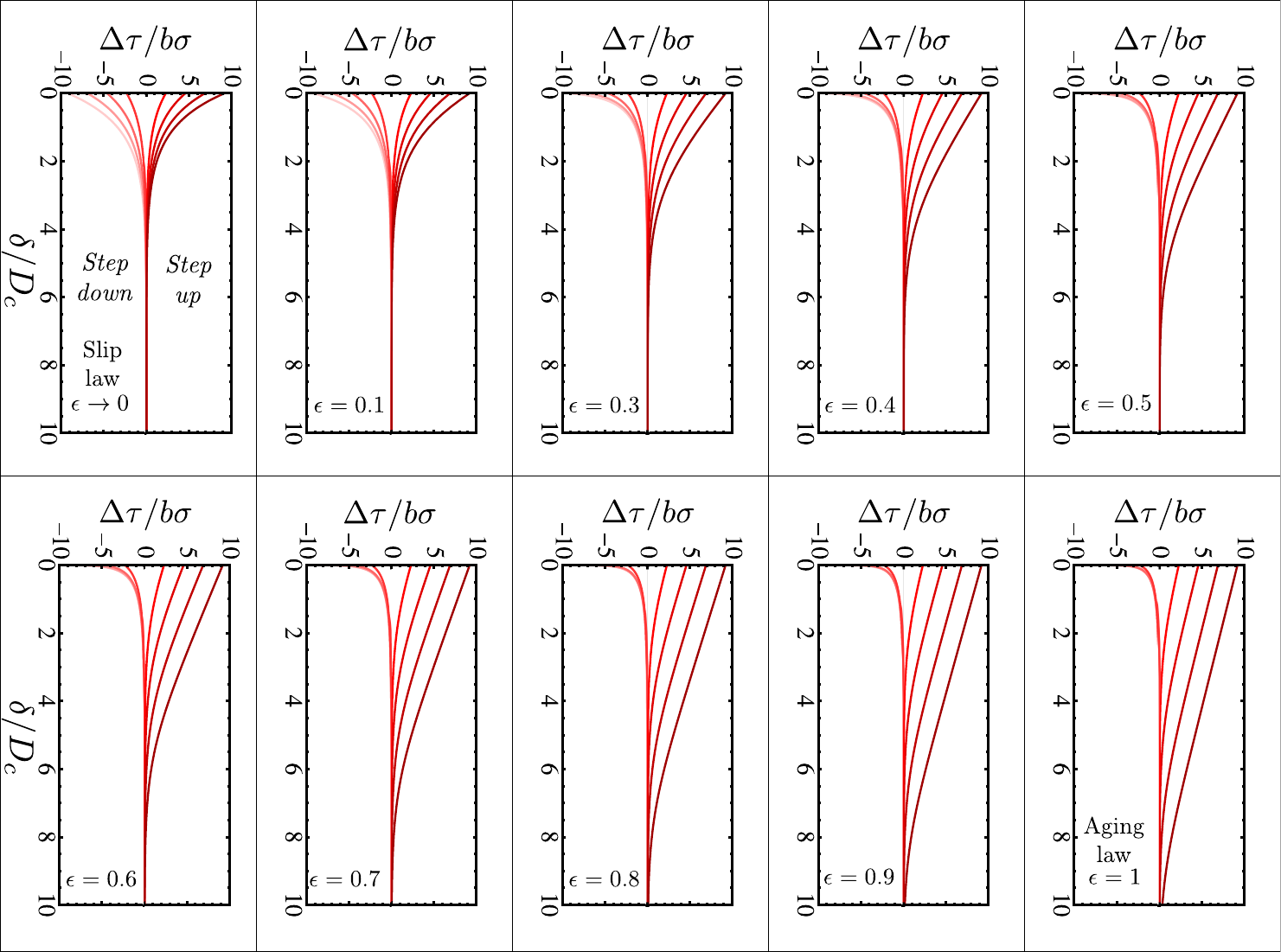}
\caption{Frictional response of spring-slider elastic system to sudden slip velocity increases (dark red) and decreases (light red) for a wide range of $V_2/V_1 = 10^{\pm (1-4)}$, where $V_1$ and $V_2$ are respectively the sliding velocity before and after the jump. The different panels display the analytical change in shear stress $\Delta \tau/b \sigma$ in function of shear displacement $\delta/D_c$ since the velocity jump (see Eq. \ref{eq: delta tau}), for different values of frictional state parameter $\epsilon$, ranging from $\epsilon \to 0$ (slip law, top-left) to $\epsilon = 1$ (aging law, bottom-right).}
\label{fig: velocity stepping experiments}
\end{figure}

As stated beforehand, the frictional state variable $\theta$ evolves following the evolution law (\ref{eq: intermediate state evolution law}). The dimensionless parameter $\epsilon$ ranges between $\left[0,1\right]$: i.e. between the slip and aging-law end-members.
Figure \ref{fig: velocity stepping experiments} shows the analytical prediction of change of frictional stress in response to sudden increases or decreases of the sliding rate, for different values of $\epsilon$. The changes are calculated for a block sliding at velocity $V_1$ at steady-state $\theta_1 = D_c/V_1$, until a sudden jump in velocity $V_2$ is imposed and eventually the new steady-state value $\theta_2 = D_c/V_2$ is achieved. In this case, Equation (\ref{eq: intermediate state evolution law}) can be integrated analytically and the resulting expression can be substituted into Eq. (\ref{eq: MC criterion}) to obtain the normalised change of frictional stress $\Delta \tau / b \sigma = \left( \tau - \tau_2\right)/ b \sigma$
between two consecutive steady states:
\begin{equation}
\Delta \tau / b \sigma  = \ln \left[\left( 1 + \left( \left(\theta_1/ \theta_2\right)^{\epsilon} - 1\right) e^{-\delta/D_c}\right)^ {1/\epsilon}\right],
\label{eq: delta tau}
\end{equation}
where $\delta$ is the slip accumulated since the velocity jump. As displayed in Figure \ref{fig: velocity stepping experiments} for a range of velocity jumps $V_2/V_1 = 10^{\pm (1-4)}$, a clear asymmetric behavior with the sign of the velocity jump is obtained for large values of $\epsilon$ parameter, with a slip length-scale over which stress evolves to a new steady state that depends on the magnitude of the jump itself (as predicted by aging law, yet inconsistent with laboratory friction experiments \citep{TuWe86, BlaMa98, Ma98}).
On the contrary, low values of $\epsilon$ parameter lead to a more symmetric frictional response and a velocity jump-independent evolution length-scale. 

At steady state, the block moves at the same speed of that of the load point, and
\begin{equation}
V_{ss} = V_o, \quad \tau_{ss} = f_{ss} \sigma =  \tau_* - \sigma \left( b-a\right) \ln \left( \frac{V_o}{V_*}\right), 
\label{eq: steady state}
\end{equation}
where $\tau_* = f_* \sigma$ and the subscript $_{ss}$ refers to steady state. If $b>a$, the logarithmic dependence of the steady state shear stress (or friction coefficient) on the slip rate is negative \ $ \left( d \tau_{ss}/d V < 0 \right)$. Hence, an increase in slip rate leads to a negative change of steady state shear stress after sufficiently slip accumulation, a condition that is commonly referred as \textit{rate-weakening} condition. Vice-versa, a \textit{rate-strengthening} interface experiences a positive change of steady-state shear stress upon an increase in slip rate. Since rate-strengthening interfaces are unconditionally stable, in this contribution we restrict ourselves to rate-weakening conditions, i.e. $b/a$ is always greater than 1.

Equations (\ref{eq: rate and state friction}), (\ref{eq: intermediate state evolution law}) and (\ref{eq: MC criterion}-\ref{eq: equation of motion2}) comprise a non-linear system governing the motion of the slider. Before studying the stability of such a motion in response to perturbations, we introduce characteristic scales and non-dimensionalize the governing equations to identify the key parameters that govern the motion.
Similar to \citet{RaRi99}, we introduce the dimensionless quantities:
\begin{align}
\kappa = \frac{k D_c}{a \sigma}, \quad \psi = \frac{\tau-\tau_*}{a \sigma}, \quad  \phi = \ln \left( \frac{V}{V_*}\right), \quad v_o = \frac{V_o}{V_*}, \quad T = \frac{V_* t}{D_c}
\label{eq: characteristic scales}
\end{align}

We then combine equations (\ref{eq: rate and state friction}) and (\ref{eq: MC criterion}) with (\ref{eq: equation of motion2}) and (\ref{eq: intermediate state evolution law}) to eliminate the state variable $\theta$ and using the characteristic scales in (\ref{eq: characteristic scales}), we arrive to the dimensionless autonomous system of non-linear, first-order differential equations 
\begin{align}
&\frac{d \psi}{d T} = \kappa \left( v_o - e^{\phi} \right) \label{eq:psi_T}\\
&\frac{d \phi}{d T} = \kappa \left( v_o - e^{\phi} \right) - \frac{b/a}{\epsilon} \left[ \left( e^{\phi} \cdot e^{\left( \psi -\phi \right)/\left( b/a\right)} \right)^{-\epsilon} -1 \right] e^{\phi} \label{eq:phi_T}
\end{align}
where the evolution of both normalized shear stress $\psi$ (or friction coefficient) and logarithmic of slip velocity $\phi$ (relative to the reference value $V_*$) is only function of the following dimensionless parameters: spring stiffness $\kappa$, rate-weakening ratio $b/a$, point loading rate $v_o$ and finally the frictional state evolution parameter $\epsilon$. 

At steady state, equation (\ref{eq: steady state}) can be rewritten in dimensionless form as
\begin{equation}
\psi_{ss} = \phi_{ss} \left( 1- b/a \right), \quad \phi_{ss} = \ln \left( v_o\right)
\label{eq: dimensionless steady state}
\end{equation} 

\section{Non-linear stability analysis with different frictional evolution laws}
\label{sec: non-linear stability analysis}
Following the work of \citet{GuRi84} and \citet{RaRi99} for the slip and aging laws, we commence the non-linear stability analysis in the phase plane $\left( \phi,\psi\right)$ by eliminating $T$ from Equations (\ref{eq:psi_T}) and (\ref{eq:phi_T}) and write 
\begin{equation}
P\left( \phi, \psi \right) \text{d}\psi + Q\left(\phi, \psi \right) \text{d} \phi = 0,
\label{eq: differential form}
\end{equation} 
where 
\begin{align}
& P\left( \phi, \psi \right) = \kappa \left( v_o - e^{\phi} \right) - \frac{b/a}{\epsilon}  \left[ \left( e^ \phi \cdot e^{\left( \psi - \phi \right)/\left( b/a\right)} \right)^ {-\epsilon} - 1\right] e^{\phi} \label{eq: P equation}\\
&Q\left( \phi, \psi \right) = - \kappa \left( v_o - e^{\phi} \right) \label{eq: Q equation}
\end{align}
The differential form in (\ref{eq: differential form}) is imperfect, but we may seek a perfect differential $U$ by introducing an integrating factor $e^{q\left( \phi, \psi\right)}$, where $q\left( \phi, \psi \right)$ remains to be determined. In this case, requiring
\begin{equation}
d U = \left[ P\left( \phi, \psi \right) \text{d}\psi + Q \left( \phi, \psi \right) \text{d}\phi \right] e^{q\left( \phi, \psi \right)}
\label{eq: dU equation}
\end{equation}
to be exact implies $\dfrac{\partial U}{\partial \phi \partial \psi} = \dfrac{\partial U}{\partial \psi \partial \phi}$, hence
\begin{equation}
\frac{\partial \left[ e^{q\left( \phi, \psi \right)} P\left( \phi, \psi \right) \right]}{\partial \phi} = \frac{\partial \left[ e^{q\left( \phi, \psi \right)} Q\left( \phi, \psi \right) \right]}{\partial \psi}
\label{eq: perfect differential condition}
\end{equation}
Substituting Equations (\ref{eq: P equation}) and (\ref{eq: Q equation}) into (\ref{eq: perfect differential condition}) we obtain the equation that $q$ must satisfy
\begin{multline}
\kappa v_o \left[ \frac{\partial q\left( \phi, \psi \right)}{\partial \phi}+ \frac{\partial q\left( \phi, \psi \right)}{\partial \psi} \right]+ \frac{e^{\phi}}{\epsilon}  e^{\epsilon \cdot \frac{(1 - b/a) \phi - \psi }{b/a}} \times \\
\left[ \frac{b}{a} \left( \epsilon -1 \right) - \epsilon - \frac{b}{a} \frac{\partial q\left( \phi, \psi \right)}{\partial \phi} + e^{\epsilon \cdot \frac{(b/a -1) \phi +\psi }{b/a}} \left( b/a - \kappa \epsilon \right) \left( 1 + \frac{\partial q\left( \phi, \psi \right)}{\partial \phi} - \kappa \epsilon \frac{\partial q\left( \phi, \psi \right)}{\partial \psi} \right) \right] = 0,
\label{eq: perfect differential condition2}
\end{multline}
which does not lend itself to a general solution. However, for some particular cases, $q \left( \phi, \psi\right)$ can be readily obtained and hence the function $U$. Contours of $U = \text{const.}$ provide solutions to the system since $dU = \dfrac{\partial U}{\partial \psi} d\psi + \dfrac{\partial U}{\partial \phi} d\phi = 0$, and tangents to the contours are $\left.\dfrac{\partial \psi}{\partial \phi}\right|_{U = \text{const.}} =  \dfrac{\partial U/\partial \phi}{\partial U/\partial \psi} = \dfrac{d\psi/dT}{d\phi/dT}$.
If an analytical solution cannot be found, the non-linear system of equations (\ref{eq:psi_T}-\ref{eq:phi_T}) can be solved numerically with initial conditions.

\section{Non-stationary load point (arbitrary $v_o$)}
\label{sec: non-stationary load point}

\subsection{Case of $\kappa=\kappa_{crit}$}
\label{subsec: k = k critical}

We start by investigating the non-linear stability of slip for the particular case when the load point is non-stationary ($v_o \ne 0$) and the spring stiffness $\kappa$ is equal to the critical value obtained from linear stability analysis (\ref{eq: critical stiffness}), which in dimensionless form reads
\begin{equation}
\kappa_{crit} = b/a -1
\label{eq: dimensionless critical stiffness}
\end{equation} 
Owing to the mapping between spring stiffness and an equivalent stiffness of a continuum fault discussed in sub-section \ref{subsec: analogy spring-block continuum}, we are considering either that the dimension of a finite fault is equal to its critical value predicted by linear stability analysis, or, for an unbounded fault, that the wavelength of the applied perturbation is equal to its critical value. \citet{Ruin83, RiRu83} showed that when $\kappa = \kappa_{crit}$ infinitesimal perturbations to steady-state sliding lead to a stable and periodic slip motion with both slip and aging law: i.e., harmonic oscillations of frictional stress and sliding velocity, as shown for example in Figure \ref{fig: periodic motion} of Appendix \ref{app: appendix}. In the non-linear regime, however, the elasto-frictional response is more complex and such a well-defined critical stiffness may not exist: systems may be linearly stable, but non-linearly unstable, and the critical stiffness may depend upon perturbation amplitude and type. In the following, we will derive analytical insights on stability and quantify critical finite perturbations that initiate unstable sliding as function of the dimensionless problem parameters.


When $\kappa=\kappa_{crit}$ and $v_o\ne 0$, it can be shown that the solution to Equation (\ref{eq: perfect differential condition2}) is
\begin{equation}
q\left( \phi, \psi \right) = (\psi - \phi) \left( 1 - \epsilon \left( 1 - a/b\right)\right)
\label{eq: q function1}
\end{equation}
which can be used to integrate (\ref{eq: dU equation}) and obtain the trajectories in the phase plane $\left(\phi, \psi \right)$ in the form
\begin{equation}
U =   \frac{b}{a}\cdot  e^{q\left( \phi, \psi\right)} \cdot \left[ \frac{\kappa v_o}{(b/a) \left( 1-\epsilon \right) +\epsilon } +  \frac{e^{\phi}}{\epsilon} \left(  \frac{e^{\epsilon \cdot \frac{(1-b/a) \phi - \psi}{b/a}}}{\epsilon-1} + \frac{b/a - \kappa \epsilon}{(b/a) \left( 1-\epsilon \right) +\epsilon}  \right) \right] = \text{const.}
\label{eq: trajectories 1}
\end{equation}

\begin{figure}[t!]
\centering
\captionsetup[subfloat]{labelfont=normalsize,textfont=normalsize}
   \subfloat[$\epsilon \to 0$ (Slip law)]{
      \includegraphics[width=.4\textwidth]{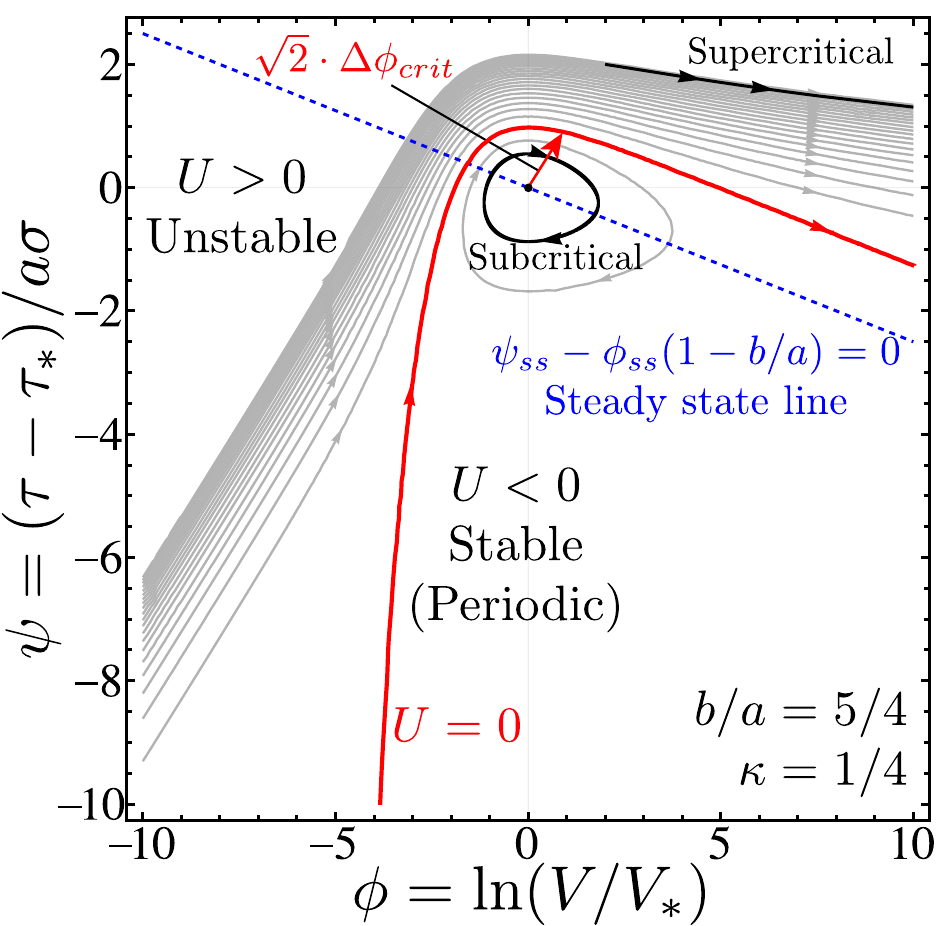}}
~
   \subfloat[$\epsilon = 0.3$]{
      \includegraphics[width=.4\textwidth]{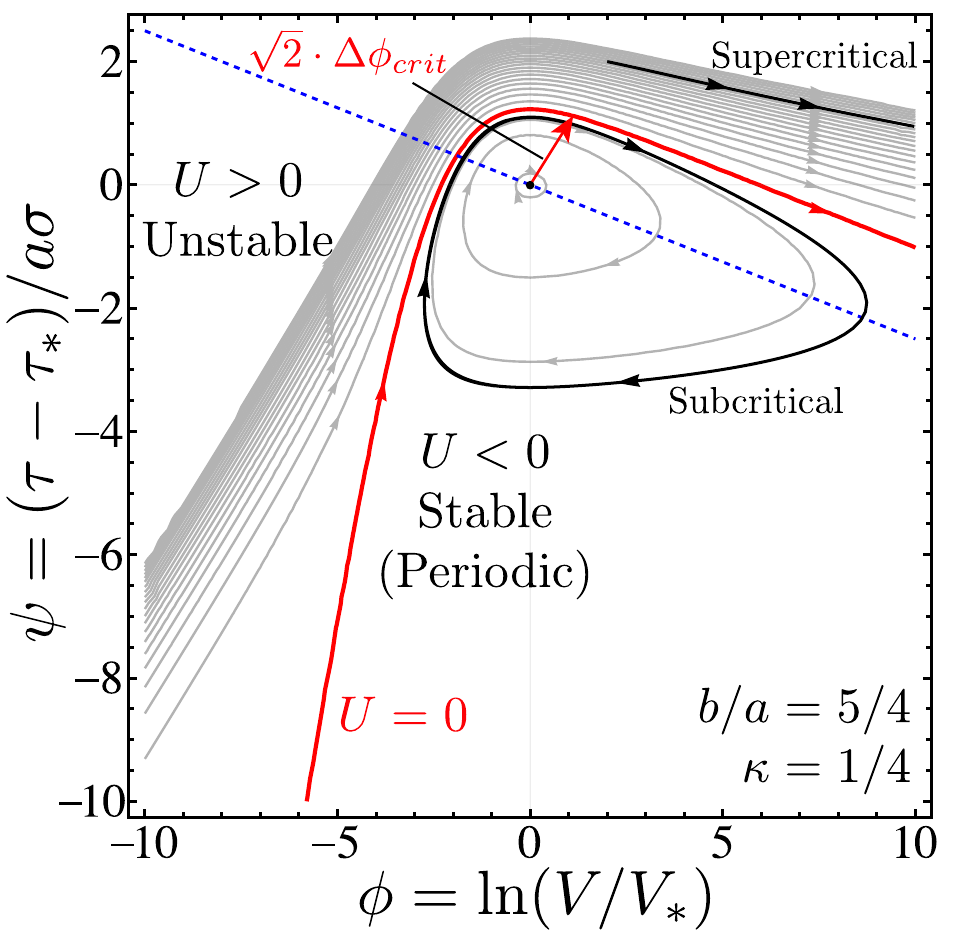}}
\\
   \subfloat[$\epsilon = 0.6$]{
      \includegraphics[width=.4\textwidth]{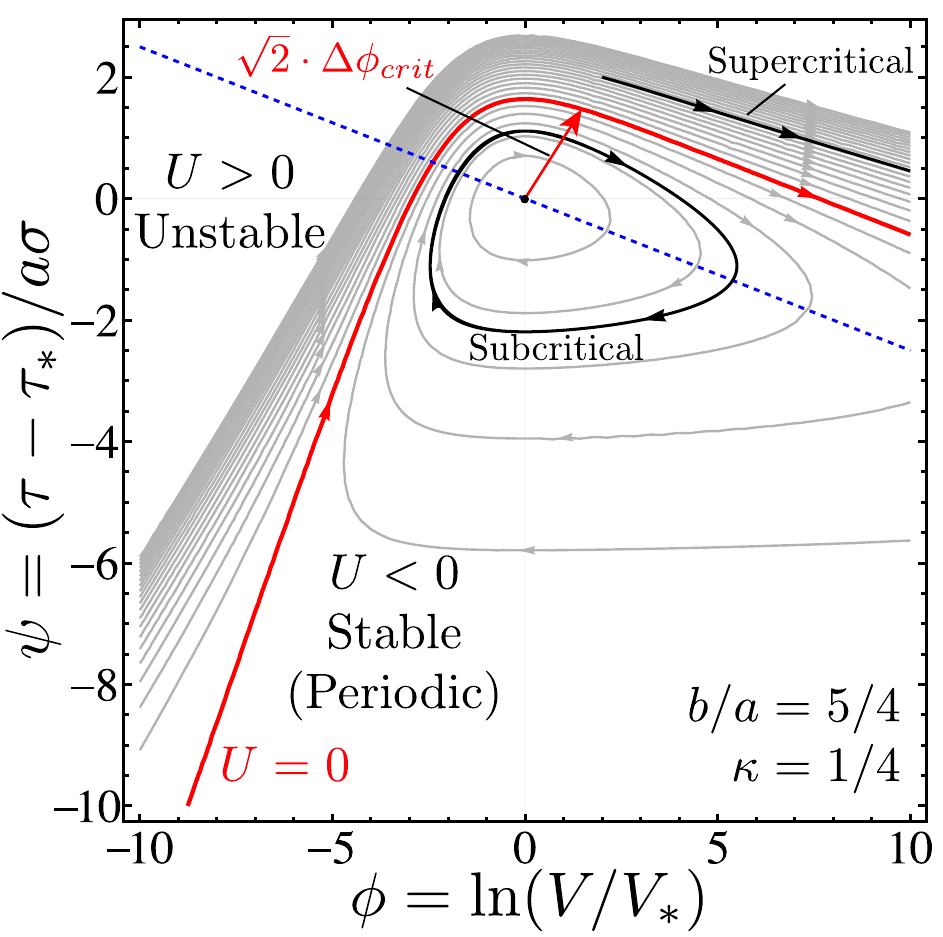}}
~
   \subfloat[$\epsilon = 1$ (Aging law)]{
      \includegraphics[width=.4\textwidth]{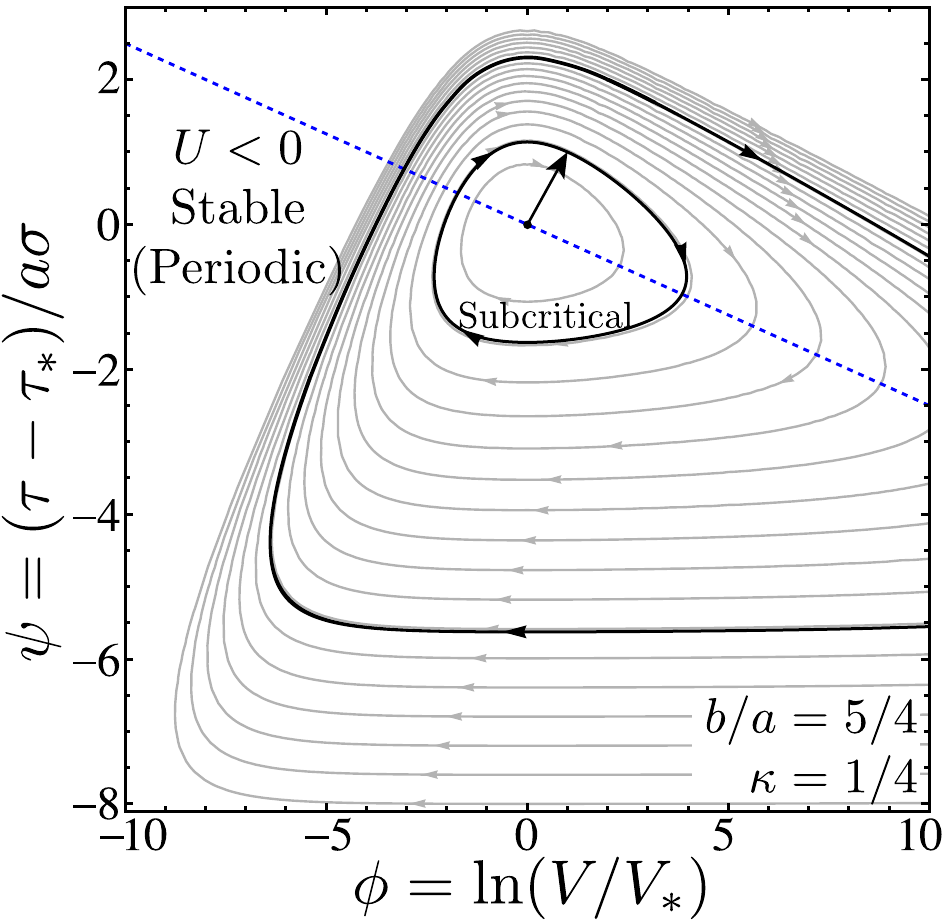}}
   \caption{Trajectories of $U$ = const. in the phase plane $\left( \phi,\psi\right)$  (see Equations \ref{eq: trajectories 1}-\ref{eq: trajectories 1-2}), for a non-stationary load point $(v_o = 1)$ and spring stiffness $\kappa$ equal to the critical value $\kappa_{crit}$. Panels (a-d) display trajectories for different values of state evolution parameter $\epsilon$, a specific value of rate-weakening ratio $b/a = 5/4$ and dimensionless spring stiffness $\kappa = 1/4 = \kappa_{crit}$. The panel (a) displays trajectories obtained with the slip law ($\epsilon \to 0$), while panel (d) with the aging law ($\epsilon = 1$). Panels (b) and (c), display trajectories for two intermediate state evolution laws, respectively $\epsilon = 0.3$ and $0.6$. Finally, on each panel the blue dashed line indicate the steady-state line, the solid red line represents the critical trajectory that separates stable (periodic) and unstable trajectories, and the black solid lines are trajectories obtained by numerically solving the non-linear system of first-order differential equations (\ref{eq:psi_T}-\ref{eq:phi_T}) with initial conditions consistent with an initial velocity jump.}
   \label{fig: trajectories 1}
\end{figure}
Note that the analytical trajectories (\ref{eq: trajectories 1}) are valid in the slip law limit as well as for any intermediate state evolution law, except in the aging law limit for which a singularity appears. Nevertheless, one can integrate (\ref{eq: dU equation}) using the specific solution (\ref{eq: q function1}) with $\epsilon = 1$ and retrieve the aging-law trajectories of \citet{RaRi99} in the form 
\begin{equation}
U =  \frac{b}{a} \left[e^{\left( \psi - \phi\right)/(b/a)} \left( v_o + \frac{e^{\phi}}{b/a - 1}\right) \kappa - \psi \right] = \text{const.},
\label{eq: trajectories 1-2}
\end{equation}
Figure \ref{fig: trajectories 1} displays examples of constant-$U$ trajectories in the phase plane $\left( \phi, \psi\right)$, for a given value of rate-weakening ratio $b/a = 5/4$, spring stiffness $\kappa = 1/4 = \kappa_{crit}$ and different values of $\epsilon$. 

While the trajectories provide a complete solution, we may go on to focus on particular scenarios. For example, consider a block initially sliding at steady-state with the same velocity as the load point, 
 i.e. $V(0)=V_o$ and hence $\phi(0) = \ln \left( V_o/V_*\right)$ and $\psi(0) = \phi \left(1-b/a\right)$. The initial state of the system in the phase plane reduces to a point on the steady-state line defined by Equation (\ref{eq: dimensionless steady state}) and depicted by blue dashed lines in Figure \ref{fig: trajectories 1}. 
Since the reference constant velocity $V_*$ may be chosen arbitrary, without loss of generality here we assume that $V_*$ is equal to the initial loading point velocity $V_o$ (i.e. $v_o = 1$), implying that the initial steady-state in the phase plane finally reduces to the origin point $(\phi(0) = 0, \psi(0) =0)$.
 When a perturbation in the form of an increase of slip rate that occurs too quickly for appreciable state evolution is introduced, the corresponding perturbation in the phase plane has changes of $\psi$ and $\phi$ of the same amount ($\Delta \phi = \Delta \psi$), owing to the constant frictional state during the velocity jump 
Such a jump follows a direction in the phase plane oriented at $45^{\circ}$ from the horizontal.

From panels (a-d) in Fig. \ref{fig: trajectories 1} we can observe that when sufficiently small perturbations to steady state are applied to the system, the state in the phase plane always jumps to a closed orbit, regardless of the value of the state evolution parameter $\epsilon$. Closed orbits in the phase plane are stable and characterized by $U<0$. To prove this, let us consider for instance orbits in the slip law limit. For a given value of $U<0$ and $\psi$ (positive or negative), Equation (\ref{eq: trajectories 1}) in the limit of $\epsilon \to 0$ can be rewritten as
\begin{equation}
 \left( b/a - 1\right) e^{-\phi} = \text{const.} + b/a - (b/a-1) \phi,
\end{equation}
where the constant in right-hand side is always negative. From a graphical construction, it can be easily seen that there are either two or no values of $\phi$ that satisfy the above equation. Similarly, for specified values of $\phi$, $b/a$ and $U<0$, we can rewrite (\ref{eq: trajectories 1}) in the slip law limit as 
\begin{equation}
U e^{-\psi} = \text{const.} + \psi,
\end{equation}
which shows again that it can be satisfied by either two or no values of $\psi$. Since for a given value of a phase variable there exist two solutions that satisfy (\ref{eq: trajectories 1}), the corresponding orbit in the phase plane must be necessarily closed.
Using a similar approach, one can further show that this is valid for any values of $\epsilon$, including the aging law limit $\epsilon = 1$. In this limit, however, equation (\ref{eq: trajectories 1-2}) must be used (see also \citet{RaRi99} for more details in this case).
Since closed orbits are periodic (as further highlighted by the gray arrow in Fig. \ref{fig: trajectories 1} indicating the direction of periodic motion on closed orbits), a stable (bounded) and periodic slip motion is thus expected. 

If the perturbation from an initial steady-state is sufficiently large, however, the state in the phase plane may jump from the origin point to an unstable trajectory for $0\leq\epsilon<1$, i.e. an open trajectory characterized by $U>0$, for which slip velocity becomes unbounded in finite time: when $U>0$, it follows from (\ref{eq: trajectories 1}) that 
\begin{equation}
e^{\epsilon \frac{\left( 1 - b/a\right)\phi - \psi}{b/a}}  < \frac{-\kappa \epsilon \left( 1- e^{-\phi}\right) + b/a}{b/a + \epsilon / \left( 1-\epsilon\right)},
\label{eq: inequality}
\end{equation}
and, using this inequality in (\ref{eq:phi_T}), it can be shown that 
\begin{equation}
\frac{d \phi}{d T} >  e^{\phi} + \frac{\left( b/a -1\right) \epsilon}{b/a - \left( b/a - 1\right) \epsilon}
\label{eq: instability condition U > 0}
\end{equation}
\begin{figure}[t!]
\centering
   \subfloat[]{
      \includegraphics[width=.475\textwidth]{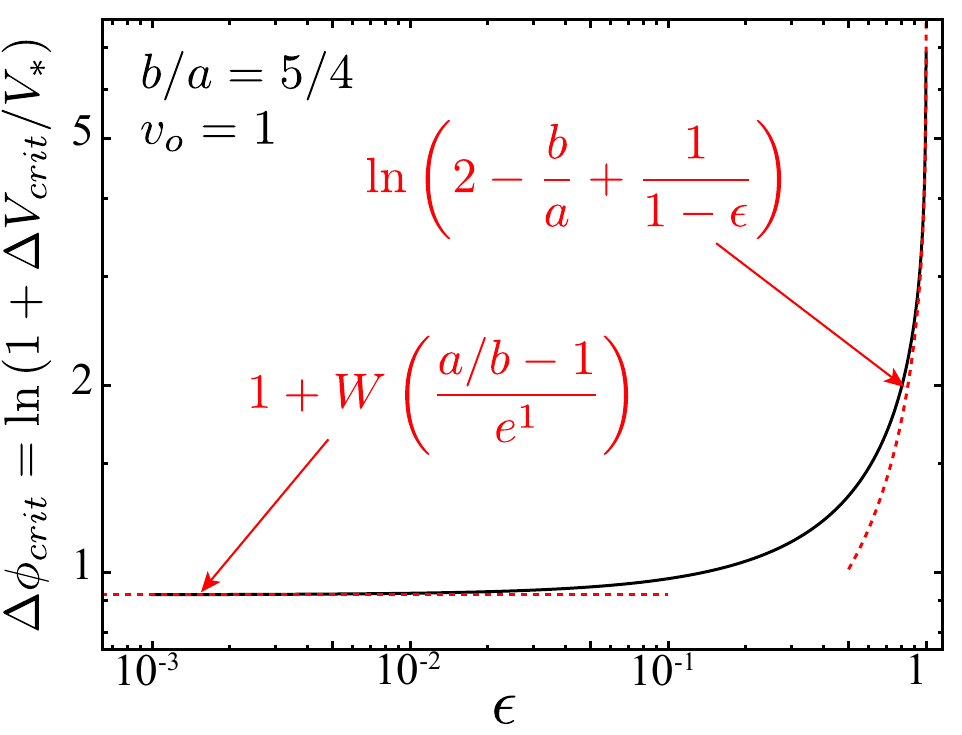}}
~
   \subfloat[]{
      \includegraphics[width=.49\textwidth]{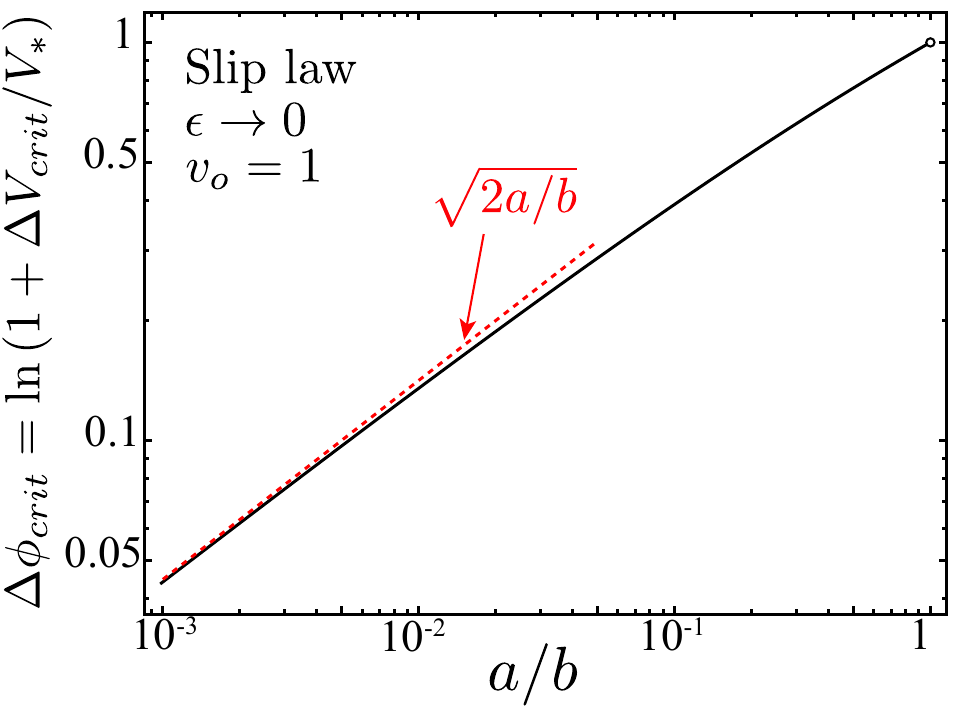}}
   \caption{Panel (a): Critical size of a sudden velocity jump $\Delta V_{crit}$ from steady-state precipitating instability, as function of $\epsilon$ and expressed as $\Delta \phi_{crit} = \ln\left( 1 + \Delta V_{crit}/V_*\right)$. The spring stiffness $\kappa = \kappa_{crit} = 1/4$ and the rate weakening ratio is $b/a = 5/4$. 
   The black solid line corresponds to the numerical solution to the system of Eq. (\ref{eq: limiting trajectory 1}) and $\psi - \phi = 0$. The two red dashed lines are the asymptotic solutions (\ref{eq: asymptotic solution critical velocity jump aging law}) and (\ref{eq: asymptotic solution critical velocity jump slip law}), for the aging-law and slip-law limits. Panel (b): Same critical jump in sliding velocity from steady state, but here as function of rate-weakening ratio $a/b$ in the slip law limit. The dashed red line defines the small $a/b$ scaling. The open circle indicates that the critical velocity jump does not exist for rate neutral frictional conditions, i.e. when $a/b = 1$.}
\label{fig: limiting perturbation 1}
\end{figure}
implies a finite-time instability, since the right-hand side of (\ref{eq: instability condition U > 0}) is strictly positive and finite for $b/a>1$ and $\epsilon \in \left[ 0,1\right]$, and since $\int_{\phi}^{\infty} e^{-\phi^{\prime}} \text{d}\phi^{\prime}$ is always bounded. Although this behavior was known to exist in the slip law limit \citep{GuRi84}, here we find it also occurs for any $\epsilon \in \left[ 0,1\right)$, excluding the aging law limit from this interval of $\epsilon$. 

When $\epsilon = 1$, indeed, the trajectories form only closed orbits in the phase plane \citep{RaRi99}, as shown in Fig. \ref{fig: trajectories 1}-d. This can be easily proved by rewriting Equation (\ref{eq: trajectories 1-2}) as \citep{RaRi99}
\begin{equation}
\text{const.} = e^{-\phi/(b/a)} + \frac{e^{\phi (b/a-1) / (b/a)}}{\left( b/a -1\right)}
\label{eq: trajectories 1-2-2}
\end{equation}
where the left-hand side is constant for given values of $U<0$, $\psi$, $b/a$, $v_o$ and $\kappa$. Therefore, like before, graphical construction of the left- and right-hand side reveals that there are either two or no values of $\phi$ that satisfy (\ref{eq: trajectories 1-2-2}). A graphical construction of the function on the right-hand side of Equation (\ref{eq: trajectories 1-2-2}) indeed resembles a parabola. When $\phi\to \pm \infty$, the value of the function is $+\infty$. The minimum of the function instead occurs for $\phi = 0$, for which the value of the function is $1+1/(b/a-1)$. Similarly, for given values of $U<0$, $\phi$ and $b/a$, Eq. (\ref{eq: trajectories 1-2}) can be re-expressed as 
\begin{equation}
U + \psi \cdot \left(b/a\right) = \text{const.} \cdot  e^{\psi/(b/a)},
\end{equation}
which is satisfied by either two or no values of $\psi$.
A dynamic instability, therefore, can never be triggered if the frictional state evolution law follows strictly the aging law:  perturbations of any size always lead to periodic oscillations, albeit with amplitudes that increase with those of the perturbations. On the contrary, when the state evolution deviates, even slightly, from the aging law, there exists a critical trajectory that separates stable (periodic) and unstable orbits, with its distance to steady state decreasing with the parameter $\epsilon$ (reaching a minimum value in the slip law limit $\epsilon \to 0$). This critical trajectory, represented in panels (a-c) of Figure (\ref{fig: trajectories 1}) by red solid lines, corresponds to $U = 0$ and is the curve given by, for $0<\epsilon<1$,
\begin{equation}
\psi_{crit} \left( \phi \right) = \frac{b/a}{\epsilon} \cdot \ln \left[ \frac{e^{\phi + \left( a/b - 1\right) \epsilon \phi } \left( b/a - \epsilon (b/a-1) \right)}{(\epsilon - 1) \left( e^{\phi} (\kappa \epsilon - b/a) - \kappa \epsilon \right)} \right],
\label{eq: limiting trajectory 1}
\end{equation}
with its expression that further simplifies in the slip law limit ($\epsilon \to 0$) as
\begin{equation}
\psi_{crit} \left( \phi \right) = 1 - \kappa \left( e^{\phi} -1 \right) - \phi (b/a-1)
\label{eq: limiting trajectory 1b}
\end{equation}
The critical trajectory is still open: given a value of $\phi$, $b/a$ and $\epsilon$ in (\ref{eq: limiting trajectory 1}), one would get only a single value of $\psi$. 

We may derive the rate of divergence, since when $U=0$, the inequalities (\ref{eq: inequality}) and (\ref{eq: instability condition U > 0}) turn into equalities. Along the critical orbit, the variation of frictional shear stress with respect to a change in slip velocity is given by 
\begin{equation}
\frac{d \psi_{crit}}{d \phi} = 1 + \frac{b}{a} \left( -1 + \frac{\kappa}{\kappa \epsilon + e^\phi \left( b/a - \kappa \epsilon\right)}\right),
\end{equation}
whose right-hand side is always negative for any value $\phi > 0$, $\epsilon<1$, $b/a>1$ and $\kappa = \kappa_{crit} = b/a-1$. Therefore, the above equation indicates that a continuous loss of frictional stress is expected as the slip rate diverges.\\  

Now, a question that arises is the following: can we determine the size of the perturbation necessary to bring a system initially, say, at steady-state, to instability? Let us take, for example, a perturbation in the form of a sudden, step change in slip velocity from an initial steady state ($V=V_o = V_*$) to a new value $V = V_o + \Delta V$ (where $\Delta V$ may be positive or negative). The change is assumed to be too rapid for any appreciable evolution of state at that instant. Can we quantify the critical jump in slip velocity $\Delta V_{crit}$ (or equivalently $\Delta \phi_{crit} = \ln (1+ \Delta V_{crit}/V_o)$) that would precipitate unstable slip? Since the jump in the phase plane occurs at constant state $\theta$, $\Delta \phi = \Delta \psi$ at the instant of perturbation, meaning the trajectory in the phase space lies along a line $\psi = \phi + c$, where $c = 0$ if initially at steady-state. Thus $\Delta V_{crit}$ is determined by the intersection of this line with the curve $\psi_{crit} \left( \phi\right)$, (\ref{eq: limiting trajectory 1}): i.e., replacing the left-hand side of (\ref{eq: limiting trajectory 1}) by $\phi + c$ and solving the implicit equation for $\phi$, the solution of which we dub $\Delta \phi_{crit} = \ln \left( 1 + \Delta V_{crit}/V_*\right)$. This solution depends on the evolution-law parameter $\epsilon$ and rate-weakening ratio $b/a$.
We solved numerically for $\Delta \phi_{crit}$ using different values of $\epsilon$ in the interval $\left[0,1\right)$, keeping the same input parameters previously reported (i.e. $v_o = 1$, $b/a = 5/4$ and $\kappa = 1/4 = \kappa_{crit}$), and limiting the solutions to those for which $\Delta V>0$ (i.e. neglecting the other branch of solutions for which velocity decreases lead to instability).
The numerical results are displayed in the log-log plot of Figure \ref{fig: limiting perturbation 1}-a (black solid line). 

\begin{figure}[t!]
\centering
      \includegraphics[width=.45 \textwidth]{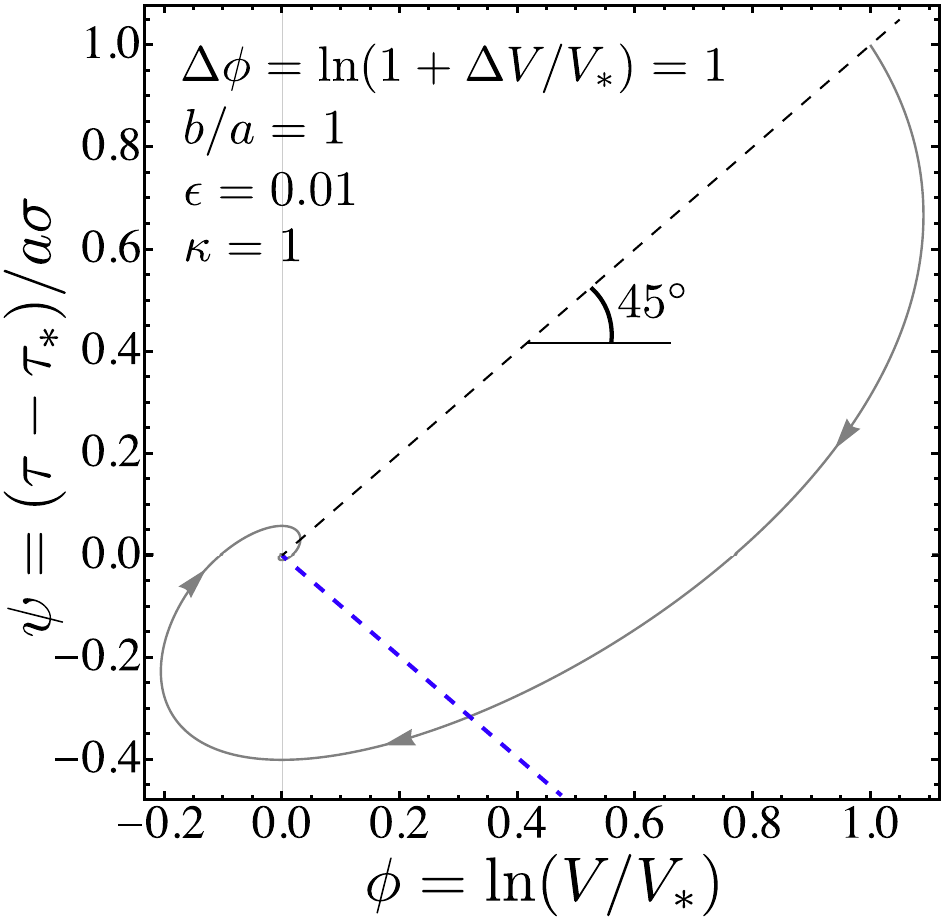}
   \caption{Example of trajectory in the phase plane when rate-neutral frictional condition is considered (i.e. $b/a=1$). The spring-slider system, initially at steady state $\left( \phi = 0, \psi = 0 \right)$ with a spring stiffness $\kappa = 1$, is subjected to a sudden perturbation in slip velocity, such to increase its initial value to $\Delta \phi = \ln(1+\Delta V /V_*) =1$ at time $T=0^{+}$). The gray trajectory define the system transient response. The blue dashed line represents the steady-state line.}
   \label{fig: rate neutral}
\end{figure}
As expected, a singularity appears when $\epsilon \to 1$: i.e. when the state tends to follow the aging (slowness) law, unconditional stability is retrieved. An asymptotic expansion of Equation (\ref{eq: trajectories 1}) allows to obtain analytical solutions in terms of critical velocity jump in the limits of $\epsilon \to 1$ (aging) and $\epsilon \to 0$ (slip). With $v_o =1$, we obtain respectively
\begin{equation}
\Delta \phi_{crit} = \ln \left( 2 - \frac{b}{a} + \frac{1}{1-\epsilon} \right)
\label{eq: asymptotic solution critical velocity jump aging law}
\end{equation}
in the aging law limit and 
\begin{equation}
\Delta \phi_{crit} = 1 + W \left( \frac{a/b-1}{e} \right)
\label{eq: asymptotic solution critical velocity jump slip law}
\end{equation}
in the slip law limit, where $W$ is the Lambert $W$-function. As shown in Figure \ref{fig: limiting perturbation 1}-a, these analytical expressions (red dashed lines) agree with the numerical solution in the two limits: a mild singularity exists as $\epsilon \to 1$ and an $\epsilon$-independent value of critical velocity jump occurs in the slip law limit to leading order. 
These results imply that a critical perturbation always exists when the frictional state evolution deviates from the aging law and $\kappa = \kappa_{crit}$. Furthermore, as $\epsilon \to 1$, $\Delta V_{crit} \sim \dfrac{1}{1-\epsilon}$ and an instability is less likely to be induced and the response to insufficient perturbation is periodic. For decreasing values of $\epsilon$, a smaller jump in velocity is needed to trigger unstable slip, with the lowest value achieved in the slip law limit and defined in (\ref{eq: asymptotic solution critical velocity jump slip law}). Additionally, in the slip law limit, the lower is the ratio $a/b$, i.e. the more pronounced is the rate-weakening condition, the lower is $\Delta \phi_{crit}$ or $\Delta V_{crit}$: the critical perturbation that triggers an instability becomes vanishingly small (see the log-log plot in Figure \ref{fig: limiting perturbation 1}-b), with an asymptotic scaling that follows $\sim \sqrt{2 \left( a/b\right)}$ in that limit (see red dashed line).
Note also that this does not apply with rate-neutral frictional condition (see exclusion circle in Fig. \ref{fig: limiting perturbation 1}-b). Indeed, when $a/b=1$, the critical spring stiffness (\ref{eq: critical stiffness}) is null, and is the spring stiffness $\kappa$ due to the assumption that $\kappa = \kappa_{crit}$. In this specific case, therefore, slip stability analysis would become irrelevant. It is worth mentioning, however, that when the spring stiffness is not equal to $ \kappa_{crit}$ and $a/b=1$, the elasto-frictional coupling emerging from spring reaction and friction at the sliding interface remains. In this case, if a perturbation in the form of an instantaneous step change in slip velocity $\Delta \phi$ is applied to the system, the frictional stress as well as the sliding velocity experience a transient evolution from their initial (and equal) values $\Delta \psi$ and $\Delta \phi$ respectively, to the values assumed before the velocity jump, i.e. the reference steady-state values $(\phi=0, \psi = 0)$. A typical response is shown in Figure \ref{fig: rate neutral} which displays the trajectory in the phase plane obtained with $b/a = 1$, $\kappa = 1 > \kappa_{crit}$, $\epsilon = 0.01$ and imposed initial velocity jump $\Delta \phi = \ln (1 + \Delta V/V_*) = 1$. The shear stress and sliding velocity spiral towards their reference steady-state values $(\phi=0, \psi = 0)$, and this is valid regardless of the perturbation strength as well as of the frictional state parameter $\epsilon$. A dynamic instability thus can never develop with rate-neutral frictional condition and $\kappa > \kappa_{crit}$ (unconditional stability). 
\subsection{Case of $\kappa \ne \kappa_{crit}$}
\label{subsec: k not = k critical}
Now, we move on to investigate the general elastic system response when the spring stiffness $\kappa$ is not necessarily equal to the critical value obtained from linear stability analysis (\ref{eq: dimensionless critical stiffness}), still keeping the point loading rate non-stationary ($v_o \neq 0$) as well as the rate weakening ratio $b/a > 1$. Under these specific conditions, we do not find the integrating factor that satisfies Equation (\ref{eq: perfect differential condition2}) analytically. Instead, we solve the non-linear system of first-order ordinary differential equations (\ref{eq:psi_T}-\ref{eq:phi_T}) numerically and investigate the stability of slip subjected to sudden perturbations in slip rate and in point loading rate.

\begin{figure}[t!]
\centering
   \subfloat[]{
      \includegraphics[width=.4\textwidth]{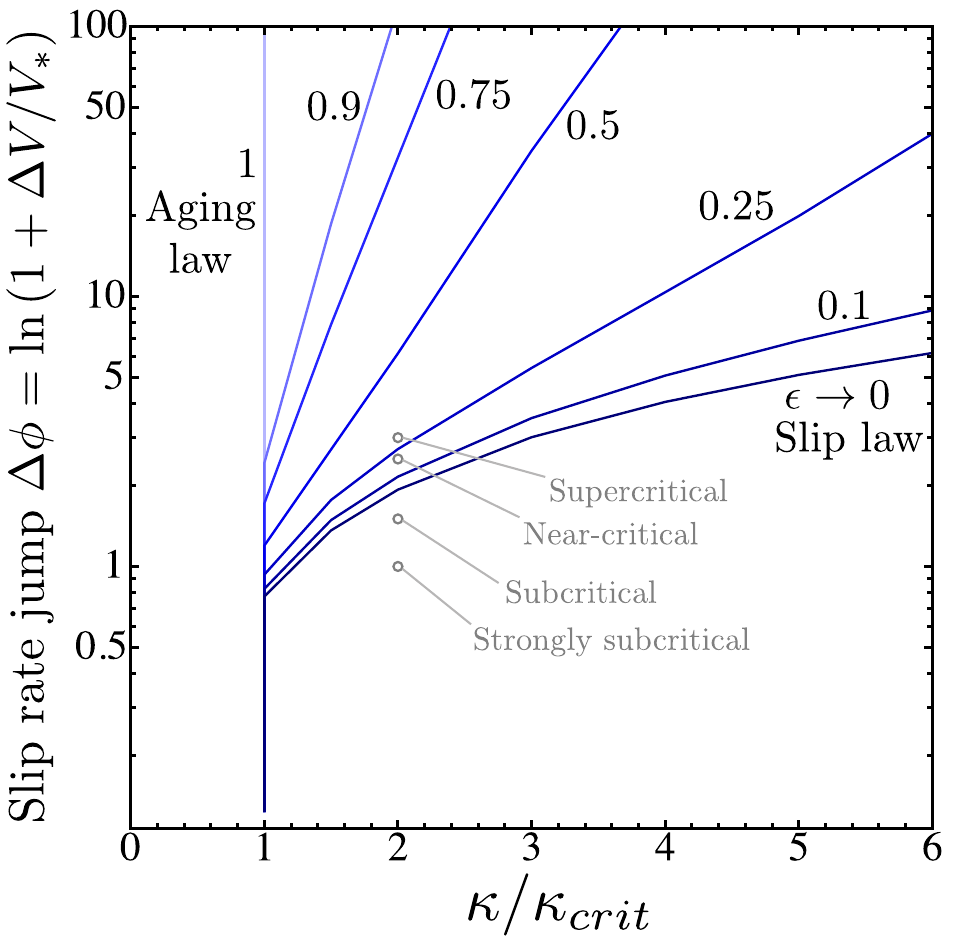}}
~
   \subfloat[]{
      \includegraphics[width=.4\textwidth]{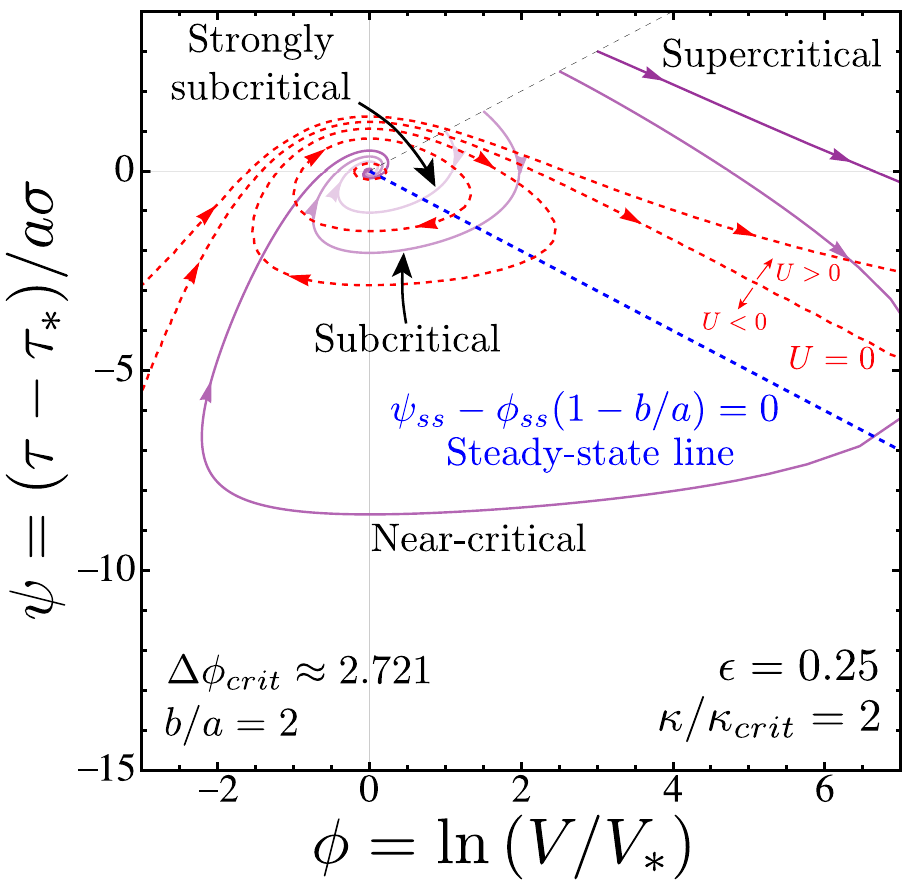}}
   \caption{Panel (a): Stability diagram in terms of log of velocity jump as a function of spring stiffness (relative to the critical value) $\kappa/\kappa_{crit}$ and frictional state parameter $\epsilon$. Each solid line represents the critical slip rate jump associated with a specific frictional state evolution law: perturbations above this line trigger a dynamic instability, whereas stable slip occurs for perturbations below this line. Light blue curves refer to aging-law-like evolution, while dark blue curves to slip-law-like evolution.
   Panel (b): solid lines are the trajectories in the phase plane $\left( \phi , \psi \right)$ associated with strongly subcritical ($\Delta \phi = 1$), subcritical ($\Delta \phi = 1.5$), near-critical ($\Delta \phi = 2.5$) and supercritical ($\Delta \phi = 3$) perturbations in slip rate. For a given value of rate weakening ratio $b/a = 2$, spring stiffness $\kappa = 2 \kappa_{crit}$ and frictional state parameter $\epsilon = 0.25$, velocity jumps approaching the critical value $\Delta \phi_{crit} = \ln \left( 1 + \Delta V_{crit}/V_*\right) \approx 2.721$ from below represent subcritical or near-critical perturbations, while a supercritical perturbation is a velocity jump larger than the critical value needed to initiate a dynamic instability (see panel (a)). Dashed red lines are the orbits (\ref{eq: trajectories 1}) obtained assuming $\kappa=1=\kappa_{crit}$ and provide level lines of a Liapunov function for this particular case.}
   \label{fig: stability diagram2}
\end{figure}

\paragraph{Perturbation in slip rate at fixed state} 
We first present the results of slip motion stability when the applied perturbation is in the form of an instantaneous jump in slip rate by an amount $\Delta V$. Similar to what discussed in section \ref{subsec: k = k critical}, the normalized slip rate $\phi$ experiences an instantaneous increase from an initial steady-state value $\phi = 0$ to a new larger value $\phi = \ln \left( 1 + \Delta V/ V_*\right)$. Since the slip rate jump occurs at fixed state, the scaled shear stress $\psi$ increases by the same amount from its initial value $\psi = 0$.
We examine the response assuming a rate-weakening ratio $b/a = 2$ and a point loading rate equal to $v_o = 1$.
The condition $\psi - \phi = \text{const.}$ provides the constraint for the initial dimensionless slip rate $\phi$ and shear stress $\psi$ at time $t = 0^+$. We integrate numerically the governing equations (\ref{eq:psi_T}-\ref{eq:phi_T}) with different values of dimensionless spring stiffness $\kappa$ and frictional state evolution parameter $\epsilon$, and solve for the evolution of $\phi$ and $\psi$ with dimensionless time $T = \dfrac{V_* t}{D_c}$ since the velocity jump. A critical jump in normalized slip rate $\Delta \phi_{crit} = \ln \left( 1+ \Delta V_{crit}/V_*\right)$ (or in shear stress) is then identified by the jump magnitude necessary for slip motion to become unstable.

The numerical results are summarized in the stability diagram reported in Figure \ref{fig: stability diagram2}-a, which shows whether slip motion is stable or unstable upon a given slip rate jump, depending on the spring stiffness ratio $\kappa / \kappa_{crit}$ and the frictional state evolution law. The blue shaded lines identify the critical jumps in slip rate for different values of $\epsilon$ parameter, with shades darkening towards the slip law limit. For a given value of $\kappa / \kappa_{crit} > 1$ and $\epsilon \in [0,1)$, perturbations to slip rate below the critical value $\Delta \phi_{crit} = \ln \left( 1+ \Delta V_{crit}/V_*\right)$ do not trigger a dynamic instability, and slip rate jumps above the critical value lead to a dynamic slip motion. In Fig. \ref{fig: stability diagram2}-a, we observe that the lower is the value of $\epsilon$, the wider is the interval of $\kappa / \kappa_{crit}$ in which a dynamic instability may be triggered (for a given slip rate perturbation). The opposite occurs when the frictional state evolution tends to the aging law. In this limit, the boundary line becomes vertical and centered at $\kappa/\kappa_{crit}=1$ \citep{RaRi99}. When $\kappa < \kappa_{crit}$, perturbations to steady-state sliding always lead to a dynamic instability, regardless of their strength. Conditional stability, instead, occurs when $\kappa \geq \kappa_{crit}$.

To support these results, we showcase examples of slider response in the phase plane when sub-critical, near-critical and supercritical perturbations are applied to the system (corresponding respectively to slip rate jumps below, near and above the critical value $\Delta \phi_{crit}$). 
We fix the ratio $\kappa / \kappa_{crit} = 2$, with $\kappa_{crit} = b/a-1 = 1$, and assume the frictional state evolves following an intermediate state evolution law with $\epsilon = 0.25$ (see Figure \ref{fig: stability diagram2}-a). Under these conditions, the critical jump in slip rate that initiates a dynamic instability is $\Delta \phi_{crit} \approx 2.721$. Figure \ref{fig: stability diagram2}-b displays the system response in the phase plane $\left(\phi,\psi \right)$. All trajectories (solid lines) start from $45^{\circ}$ direction originated from steady-state condition (see black dashed line) and, if the perturbation applied is sufficiently weak, they return to the initial steady-state condition ($\phi = 0$, $\psi = 0$) (see strongly sub-critical, sub-critical or near-critical trajectories) or may diverge to infinity (see supercritical trajectory). Steady-state creep is thus expected following a few decaying oscillations of slip rate and shear stress on sub-critical/near-critical trajectories, as opposed to supercritical trajectories where unstable slip motion is expected. Finally, we note the function $U$ obtained for $\kappa = \kappa_{crit}$ and mathematically given by (\ref{eq: trajectories 1}) can be regarded as a Liapunov function for this particular case, similarly to what \citet{GuRi84} obtained for the slip law. 
A Liapunov function in the phase plane can be used to verify the basin of an attraction of an equilibrium state of a non-linear dynamical system: it indicates how close solutions need to be to equilibrium to be drawn to it. 
As shown in Fig. \ref{fig: stability diagram2}-b, the solid trajectories crossing the red dashed orbit $U=0$ (i.e. the strongly sub-critical, sub-critical and near-critical trajectories) and hence entering in the region of $U<0$ where closed dashed orbits lie, are drawn towards the steady-state point $\left( \phi=0,\psi=0\right)$ at which $U$ reaches its minimum value. The supercritical trajectory, instead, diverges from the $U=0$ orbit and never crosses the closed curves: unstable behavior is then expected.

\paragraph{Perturbation in point loading rate}
We also investigate slip motion stability when the applied perturbation is not in the form of a slip rate jump, but is in the form of a jump in the point loading rate.
Let us assume that initially the block slides steadily at velocity $V = V_o = V_*$, leading to the following initial conditions for the variables $\phi$ and $\psi$: $\phi = 0$ and $\psi = 0$ at time $t = 0^+$. Let us impose then a sudden perturbation to the rate of the load point displacement, increasing instantaneously its initial value from $v_o = 1$ to $v_o + \Delta v_o$ at time $t = 0^+$. 
We fix the rate-weakening ratio to $b/a = 2$ and, for a given arbitrary value of $\kappa$ and $\epsilon \in \left( 0,1\right]$, we integrate numerically the governing equations  (\ref{eq:psi_T}-\ref{eq:phi_T}) and solve for the evolution of $\phi$ and $\psi$ with dimensionless time $T = \dfrac{V_* t}{D_c}$ since the jump in loading rate (that is equivalent to solve for slip accumulated $\delta/D_c$ since $t=0^+$). 
A critical loading rate jump $\left(\Delta v_{o}/v_o\right)_{crit}$ from steady-state condition is then identified, in correspondence of which slip motion becomes unstable and frictional strength rapidly decrease.

\begin{figure}[t!]
\centering
   \subfloat[]{
      \includegraphics[width=.4\textwidth]{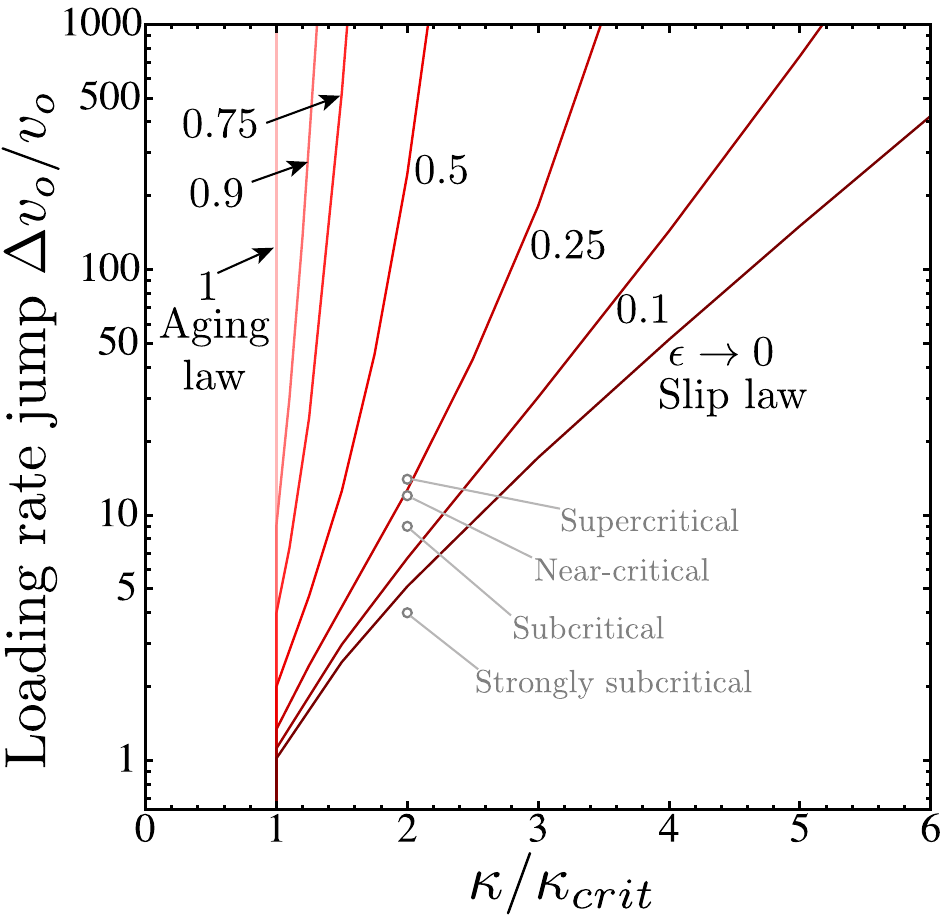}}
~
   \subfloat[]{
      \includegraphics[width=.4\textwidth]{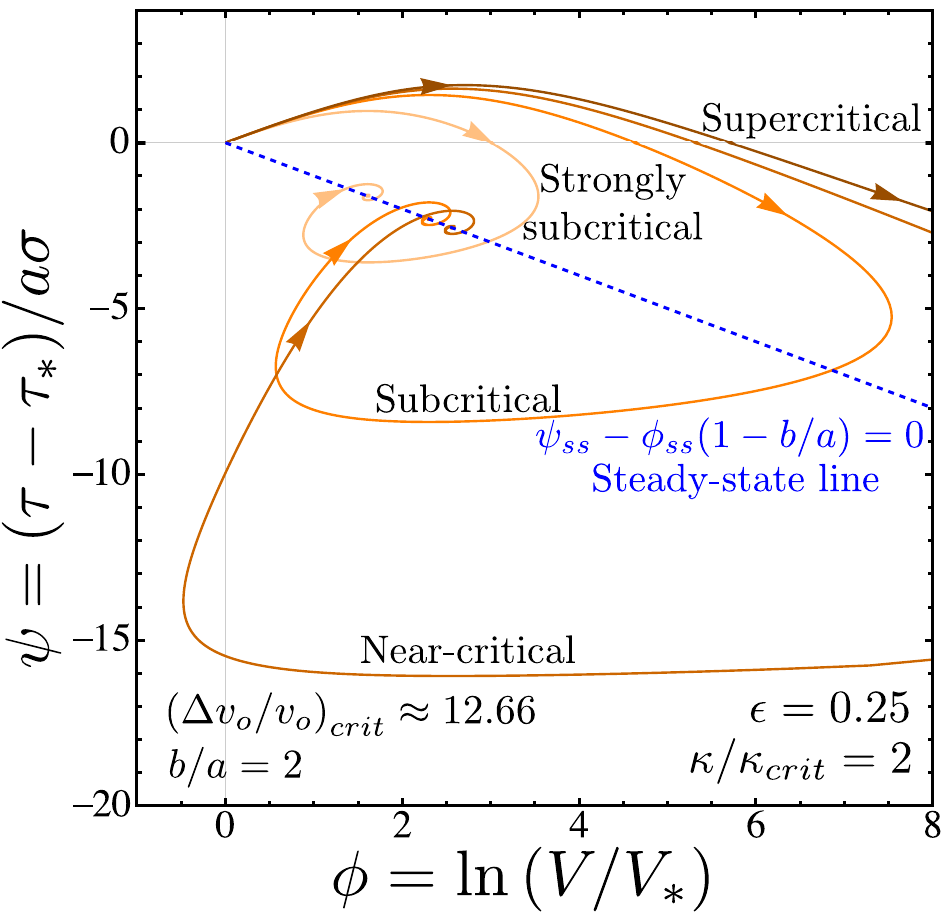}}
   \caption{Panel (a): Stability diagram in terms of loading rate jump as a function of scaled spring stiffness $\kappa/\kappa_{crit}$ and frictional state parameter $\epsilon$. Each solid line represents the critical point loading rate jump associated with a specific frictional state evolution. Light red curves refer to aging-law-like evolution, while dark red curves to slip-law-like evolution. Unstable motion occurs for perturbations to the left of a boundary, and stable to the right.
   Panel (b): trajectories in the phase plane $\left( \phi , \psi \right)$ associated with strongly subcritical ($\Delta v_o / v_o = 4$), subcritical ($\Delta v_o / v_o = 9$), near-critical ($\Delta v_o / v_o = 12$) and supercritical ($\Delta v_o / v_o = 14$) perturbations in point loading rate. For a given value of rate weakening ratio $b/a = 2$, spring stiffness $\kappa = 2 \kappa_{crit}$ and frictional state parameter $\epsilon = 0.25$, loading rate jumps approaching the critical value $\left(\Delta v_{o} / v_o\right)_{crit} \approx 12.66$ from below represent subcritical or near-critical perturbations, while a supercritical perturbation is a loading rate jump larger than the critical value needed to initiate a dynamic instability (see panel (a)).}
   \label{fig: stability diagram}
\end{figure}

The numerical results are summarized in the plots reported in Figure \ref{fig: stability diagram}. Although the stability diagram on the left-hand side looks qualitatively similar to the one obtained for perturbations in slip rate, the response of spring-slider in the phase plane differs. Plot (b) of Fig. \ref{fig: stability diagram} shows the trajectories in the phase plane obtained with the same input parameters of the previous case, i.e. $b/a =2$, $\epsilon = 0.25$ and $\kappa/\kappa_{crit} = 2$.
We observe that all the trajectories start from the initial conditions of steady-state sliding $\left( \phi = 0, \psi = 0\right)$ and, depending on the strength of the perturbation applied, the response evolves following different trajectories. Given that the change in load point velocity occurs at fixed state, all the trajectories share a common initial tangent. For loading rate jumps below or near the critical value, the trajectories spiral towards a new steady-state at a sliding rate of $v_o + \Delta v_o$ (see strongly subcritical, subcritical and near-critical trajectories in Fig. \ref{fig: stability diagram}-b).
Conversely, when the loading rate jump is above the critical value (supercritical condition), the response is characterized by a rapid and continuous loss in frictional strength and, consequently, a diverging slip rate. Slip histories of $\phi$, $\psi$ and $\theta$ for those cases are shown in Figure \ref{fig: non-periodic motion} of Appendix \ref{app: appendix}.

\section{Stationary load point ($v_o = 0$)}
\label{sec: stability stationary load point}
Here, we investigate the stability of the slider under no active loading, i.e. a stationary load point. 
We set $v_o = 0$ in Eq. (\ref{eq: perfect differential condition2}) and look for the integrating factor $e^{q\left( \phi, \psi \right)}$ that satisfies Equation (\ref{eq: perfect differential condition}) for any $\epsilon \in \left[ 0,1\right]$. It can be shown that the factor $q\left( \phi, \psi \right)$ that satisfies (\ref{eq: perfect differential condition}) is 
\begin{equation}
q\left( \phi, \psi \right) = \frac{\phi \left( b/a \left( \epsilon - 1\right) - \epsilon \right)}{b/a} + \frac{\psi \left( b/a - 1\right) \left( b/a - \kappa \epsilon\right)}{\kappa \cdot (b/a)}
\end{equation}
which can be used to integrate (\ref{eq: dU equation}) and define the corresponding trajectories in the phase plane $\left( \phi, \psi\right)$ in the form
\begin{equation}
U = \frac{  e^{\psi \left( b/a-\kappa \epsilon - 1\right) /\kappa} \cdot \dfrac{\kappa b}{a} \left( 1 - \dfrac{b}{a} +  e^{\epsilon \cdot \frac{(b/a -1) \phi +\psi }{b/a }} \left( \dfrac{b}{a} - 1 - \kappa \epsilon \right)  \right)}{\epsilon (b/a -1) (b/a - 1 -\kappa \epsilon)} = \text{const.}
\label{eq: trajectories 2}
\end{equation}
Note that, unlike the non-stationary case presented in section \ref{sec: non-stationary load point}, here the analytical trajectories represented by (\ref{eq: trajectories 2}) are valid for any intermediate state evolution law, including the end-member slip and aging laws.  

Similarly to the results of the previous section \ref{sec: non-stationary load point}, unstable orbits are those for which $U>0$. From (\ref{eq: trajectories 2}) it follows that $U>0$ when
\begin{equation}
e^{-\epsilon \cdot \left[ \left(\psi - \phi \left( 1-b/a\right) \right)/(b/a)\right]} < \frac{b/a - 1 - \kappa \epsilon}{b/a -1},
\end{equation}
which can be replaced into (\ref{eq:phi_T}) along with the condition $v_o = 0$ to obtain  
\begin{equation}
\frac{\partial \phi}{\partial T} > \frac{\kappa e^{\phi}}{b/a-1}
\label{eq: instability condition2 U > 0}
\end{equation}
Interestingly, the inequality (\ref{eq: instability condition2 U > 0}) is the same found by \citet{RaRi99} for the non-linear stability analysis of the spring-slider elastic system with the aging law (\ref{eq: aging law}) (see their Eq. (35)). 
Since the right-hand side of (\ref{eq: instability condition2 U > 0}) is strictly positive and finite for $b/a>1$ (with $\int_{\phi}^{\infty} e^{-\phi^{\prime}} \text{d}\phi^{\prime}$ bounded), then it is clear that slip velocity becomes unbounded in finite time on positive-$U$ trajectories.

Conversely, slip motion on trajectories characterized by $U<0$ is always stable. To demonstrate this expectation, let us rewrite Equation (\ref{eq: trajectories 2}) as
\begin{equation}
e^{\epsilon (b/a - 1) \phi/(b/a)} = \left( \dfrac{b}{a} - 1\right) e^{-\epsilon \psi/(b/a)} \left[ \frac{\epsilon \cdot U}{\kappa \cdot b/a} \cdot e^{(1+\kappa \epsilon -b/a) \psi/\kappa} - \frac{1}{1+ \kappa \epsilon - b/a}\right]
\label{eq: Unegative}
\end{equation}
Since the right-hand side of the above equation tends to $-\infty$ when $\psi \to -\infty$, for any $\kappa$, $\epsilon \in \left(0,1\right]$ (slip law limit excluded),  $b/a>1$ and $U<0$, there exists no value of $\phi$ that satisfies equation (\ref{eq: Unegative}). Therefore, a complete loss of frictional strength never occurs on negative-$U$ trajectories when frictional state evolution deviates from the slip law (and hence instability can never take place). Rather, $\psi$ approaches a constant value only when $\phi \to -\infty$ and $\epsilon$ does not vanish. Indeed, the left-hand side of (\ref{eq: Unegative}) vanishes when $\phi \to - \infty$ and $\epsilon >0$, and thus the limiting shear stress value achieved at very small slip velocities (for which $V \ll V_*$ or $e^{\phi}$ is much smaller than 1) is given by
\begin{equation}
\lim_{\phi \to - \infty} \psi = \left[ \frac{\kappa}{(1+\kappa \epsilon - b/a)} \cdot \ln \left(\frac{b/a}{\epsilon \cdot  U} \frac{k}{(1+\kappa \epsilon - b/a)}\right)\right],
\end{equation} 
where in the aging law limit $(\epsilon = 1)$ we retrieve the previous finding of \cite{BhaRu17}.
However, unlimited weakening does occur in the limit of small slip velocities for the slip law ($\epsilon \to 0$). To show this, Equation (\ref{eq: trajectories 2}) can be rewritten as
\begin{equation}
\left( \frac{b}{a}- 1\right)\phi = \frac{U \left( b/a - 1\right)}{\kappa} \cdot e^{-(b/a-1)\psi/\kappa}+\frac{\kappa \cdot (b/a)}{(b/a-1)} - \psi,
\label{eq: Unegative2}
\end{equation}
whose right-hand side also tends to $-\infty$ when $\psi \to -\infty$ (for $U<0$, $b/a>1$ and any $\kappa$). The only possible behavior that satisfies (\ref{eq: Unegative2}) is that for which $\phi \to - \infty$, which means that also when $\epsilon \to 0$ slip is stable, and a constant minimum value of $\psi$ is never achieved at very small velocities.

The stability boundary that divides stable and unstable motion is the trajectory for which $U=0$,
\begin{equation}
\psi_{crit}(\phi) = \left( 1 - \frac{b}{a}\right) \phi + \frac{b}{a \epsilon} \cdot \ln \left( \frac{b/a - 1}{b/a -1 - \kappa \epsilon} \right)
\label{eq: limiting trajectory2}
\end{equation}
The critical trajectories in this case are straight lines in the phase plane, with the ratio $b/a$ governing their slopes and the frictional state evolution parameter $\epsilon$ affecting solely their shift from the steady state line (\ref{eq: dimensionless steady state}).
$\psi_{crit}(\phi)$ exists if and only if
\begin{equation}
\frac{\kappa \cdot \epsilon }{\kappa_{crit}} < 1.
\label{eq: critical trajectory condition}
\end{equation}
When this condition is violated, all the trajectories tend towards $\phi \to -\infty$ and are stable. 
This suggests there exists a stiffness $\kappa$ above which an instability can never develop. This stiffness, however, increases with decreasing values of $\epsilon$, and diverges as $\epsilon \to 0$. In the slip law limit, therefore, perturbations to the elasto-frictional system may potentially initiate a dynamic instability, provided that their magnitudes drive the system to cross the threshold defined in (\ref{eq: limiting trajectory2}) for the limit $\epsilon \to 0$: $\psi_{crit} = \left(1-\dfrac{b}{a}\right) \phi + \dfrac{\kappa (b/a)}{b/a - 1}$. For a given value of $b/a$ and spring stiffness $\kappa$, the condition (\ref{eq: critical trajectory condition}) also defines a critical value of $\epsilon$ above which slip motion is unconditionally stable
\begin{equation}
\epsilon_{crit} = \frac{\kappa_{crit}}{\kappa},
\label{eq: critical epsilon}
\end{equation}
or, similarly,  a critical value of $a/b$ above which $\psi_{crit}(\phi)$ does not exist (for a given $\epsilon$ and stiffness $\kappa$), i.e. 
\begin{equation} 
(a/b)_{crit} = \left(\kappa \epsilon + 1 \right)^{-1}
\label{eq: critical ab}
\end{equation}

\begin{figure}[t!]
\centering
\captionsetup[subfloat]{labelfont=normalsize,textfont=normalsize}
   \subfloat[$\epsilon \to 0$ (Slip law)]{
      \includegraphics[width=.4\textwidth]{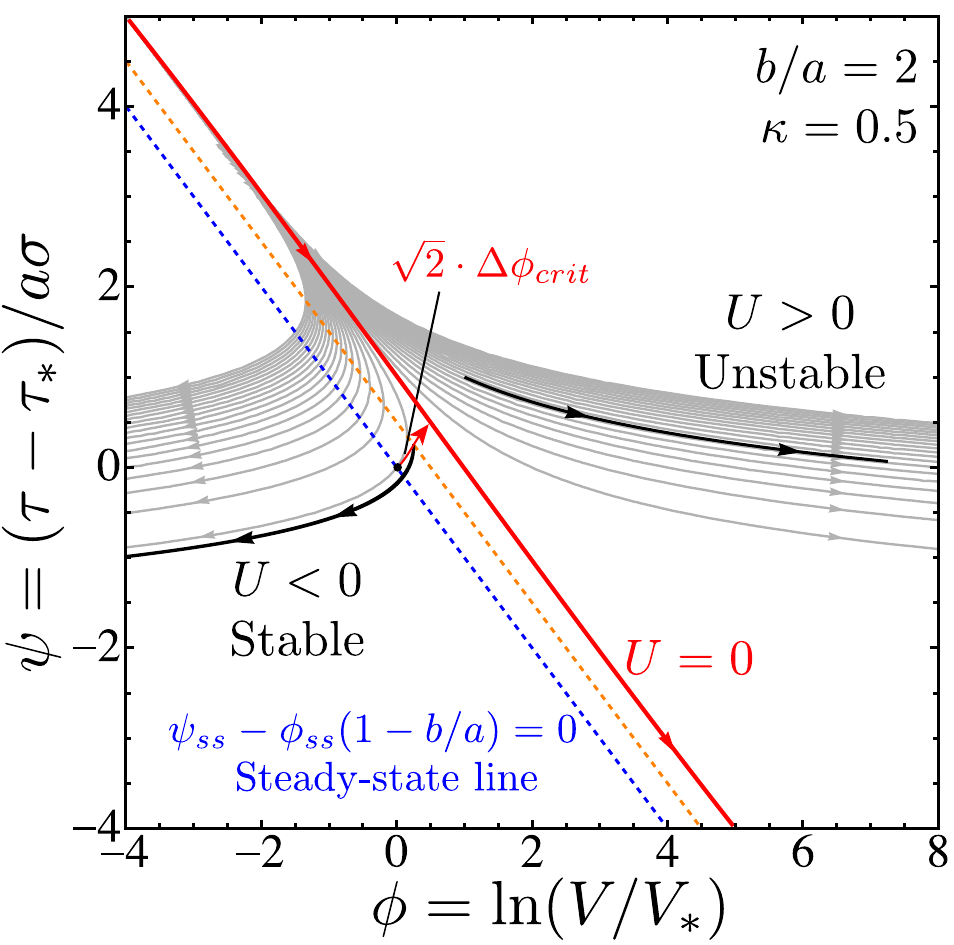}}
~
   \subfloat[$\epsilon = 0.3$]{
      \includegraphics[width=.4\textwidth]{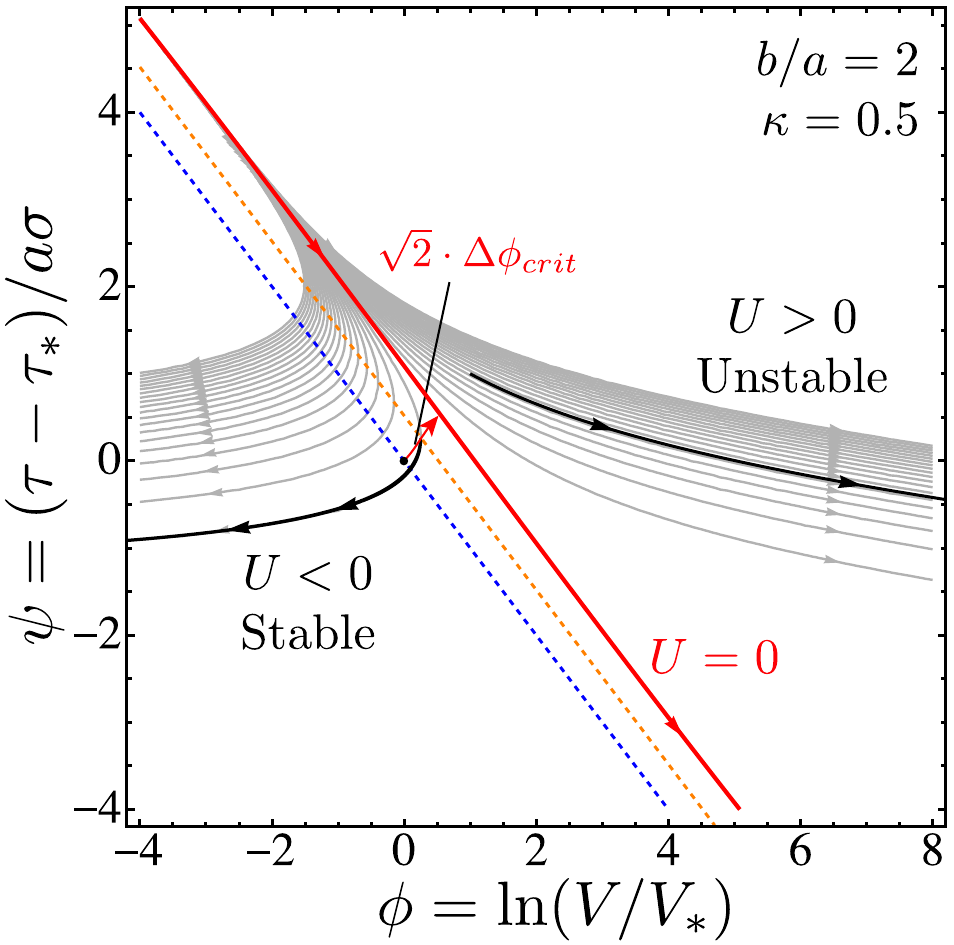}}
\\
   \subfloat[$\epsilon = 0.6$]{
      \includegraphics[width=.4\textwidth]{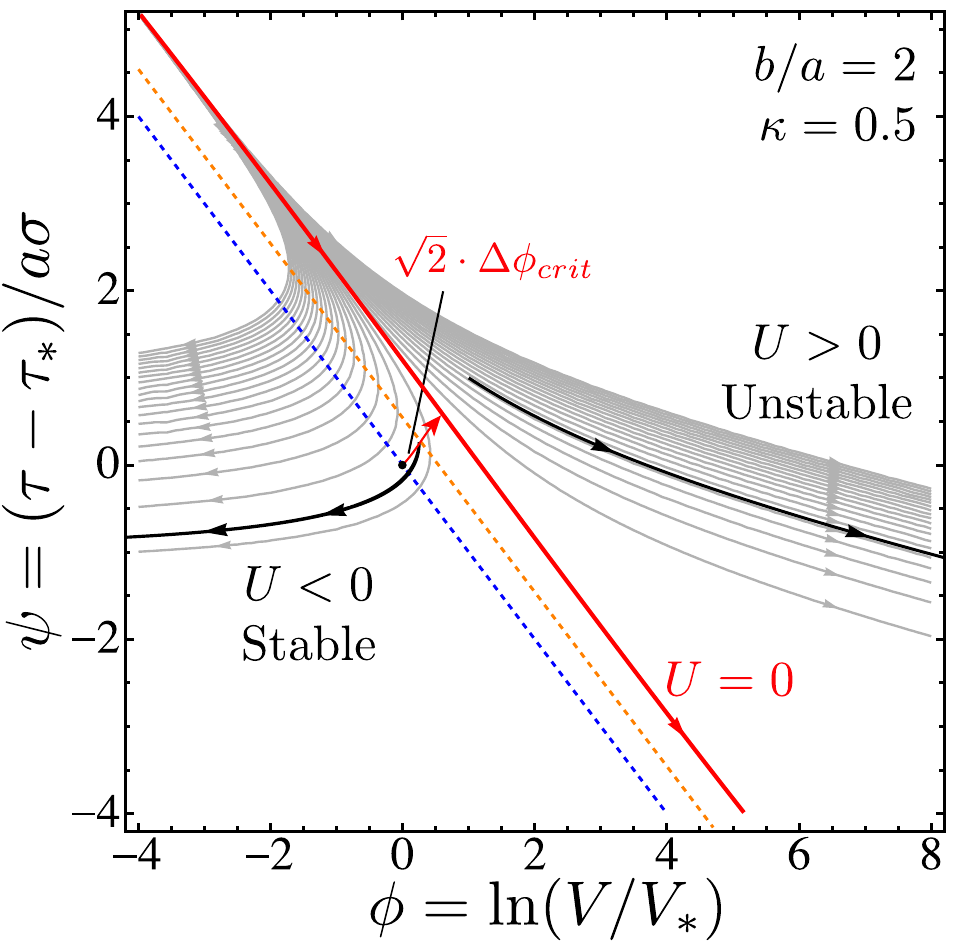}}
~
   \subfloat[$\epsilon = 1$ (Aging law)]{
      \includegraphics[width=.4\textwidth]{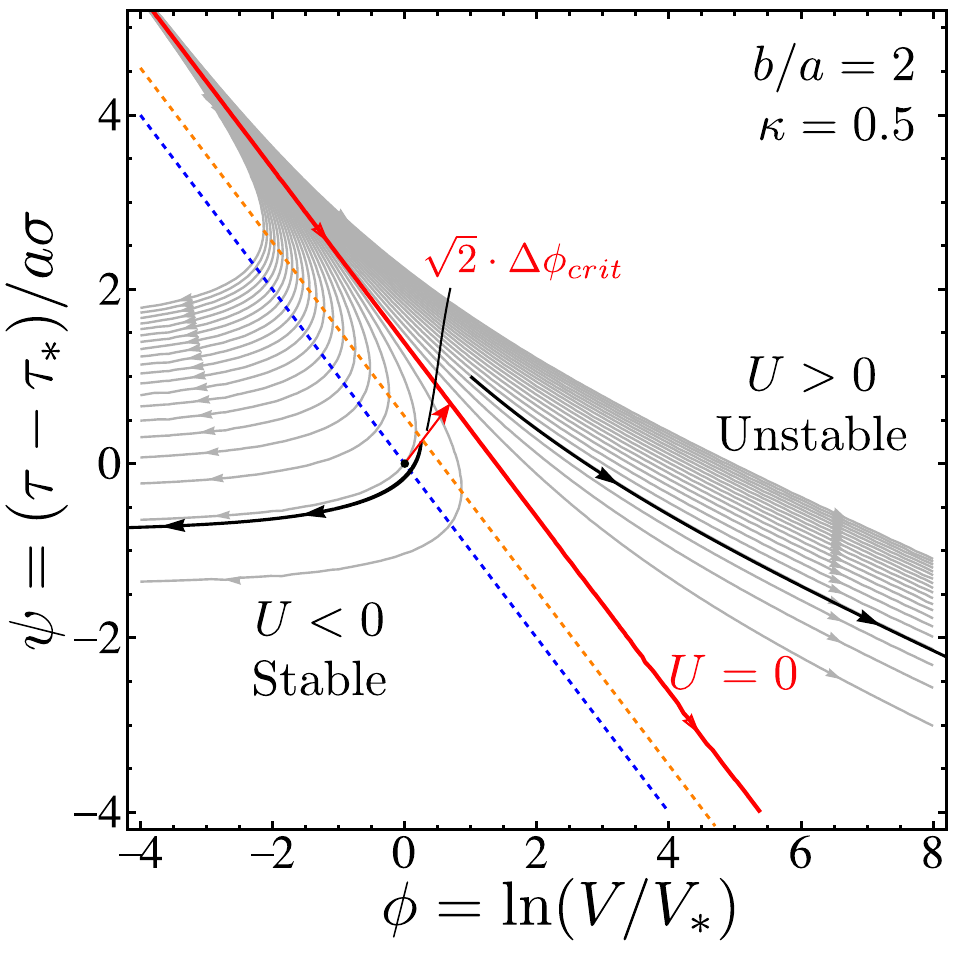}}
   \caption{Trajectories in the phase plane $\left( \phi,\psi\right)$ given by the level-set curves of Eq. (\ref{eq: trajectories 2}), for stationary load point $(v_o = 0)$ and spring stiffness $\kappa$ below the critical value $\kappa_{crit}$. Panels (a-d) display trajectories for different values of frictional state evolution parameter $\epsilon$, a specific value of rate-weakening ratio $b/a = 2$ and dimensionless spring stiffness $\kappa = 0.5 < \kappa_{crit}$. The panel (a) displays trajectories obtained with the slip law ($\epsilon \to 0$), while panel (d) with the aging law ($\epsilon = 1$). Panels (b) and (c), instead, display the elastic system response for two intermediate state evolution laws, respectively $\epsilon = 0.3$ and $0.6$. On each panel, the blue dashed line refers to steady state line, whereas the orange dashed lines refers to acceleration lines, which denote up to which point motion on an intersecting trajectory accelerates. Finally, the solid red line represents the critical trajectory that separates stable and unstable slip (see Equation \ref{eq: limiting trajectory2}), whereas the black solid lines are trajectories obtained numerically by solving the non-linear system of first-order differential equations (\ref{eq:psi_T}-\ref{eq:phi_T}) with initial conditions consistent with an initial velocity jump.}
   \label{fig: trajectories 2}
\end{figure}

\begin{figure}[t!]
\centering
\captionsetup[subfloat]{labelfont=normalsize,textfont=normalsize}
   \subfloat[$\epsilon \to 0$ (Slip law)]{
      \includegraphics[width=.4\textwidth]{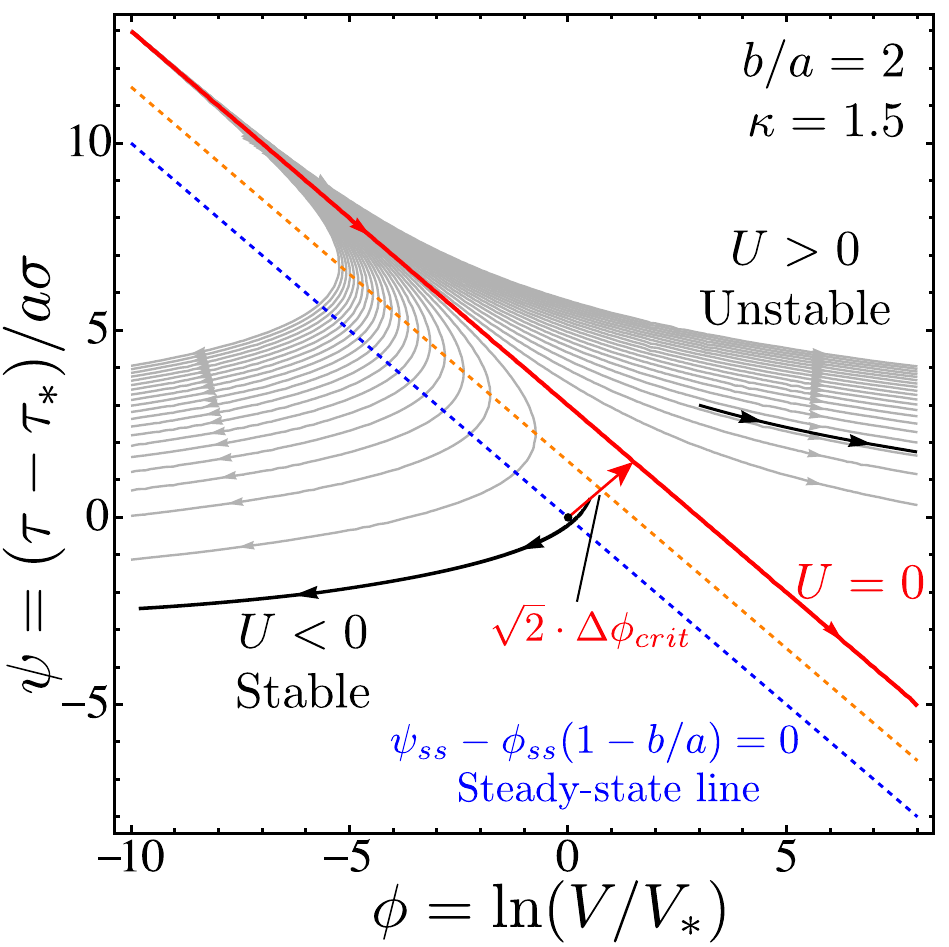}}
~
   \subfloat[$\epsilon = 0.6$]{
      \includegraphics[width=.4\textwidth]{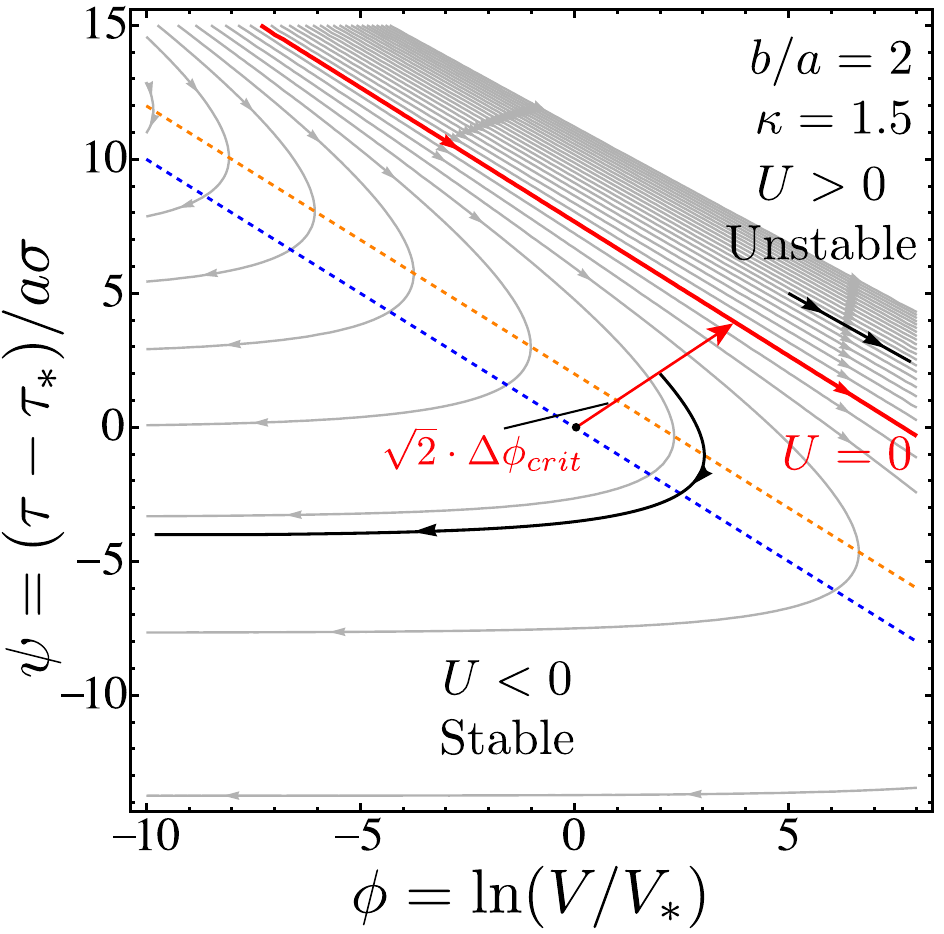}}
\\
   \subfloat[$\epsilon = 0.7$]{
      \includegraphics[width=.4\textwidth]{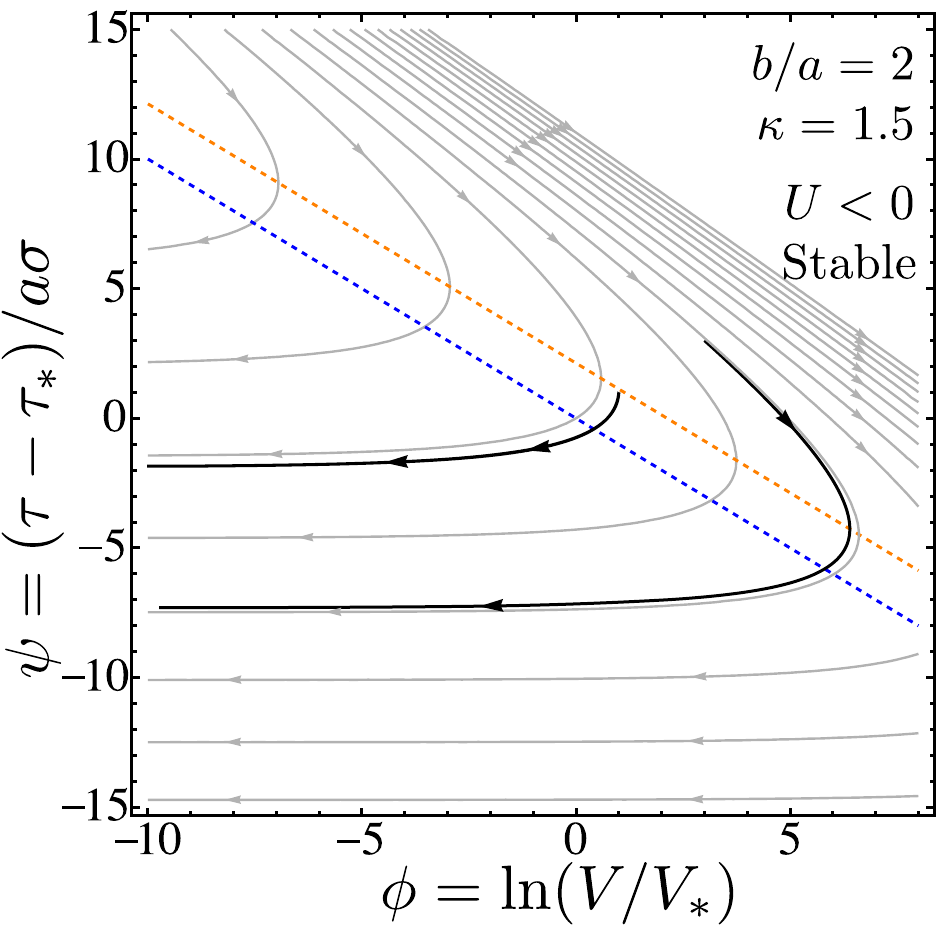}}
~
   \subfloat[$\epsilon = 1$ (Aging law)]{
      \includegraphics[width=.4\textwidth]{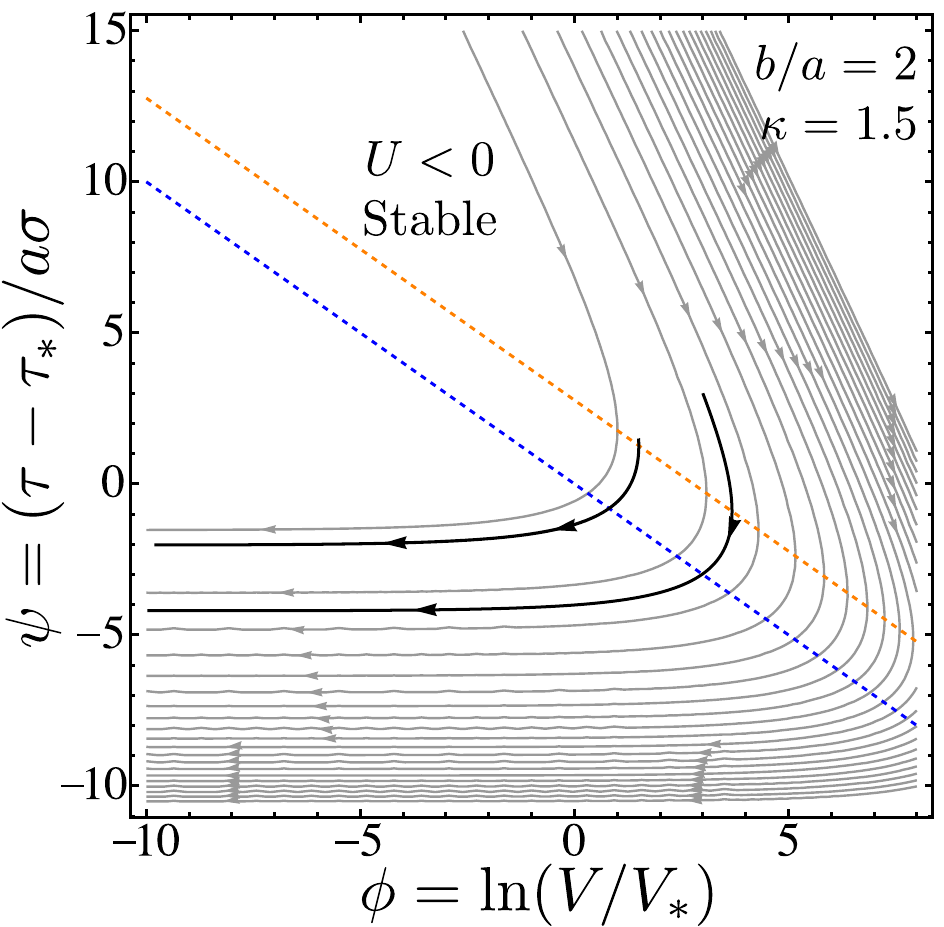}}
   \caption{Trajectories in the phase plane $\left( \phi,\psi\right)$  (given by Equation \ref{eq: trajectories 2}), for stationary load point $(v_o = 0)$ and spring stiffness $\kappa$ larger than the critical value $\kappa_{crit}$, demonstrating the vanishing of unstable motion with $\epsilon$ increasing past a critical value $\epsilon_{crit} = 2/3$ given by (\ref{eq: critical epsilon}). Panels (a-d) display trajectories for different values of frictional state evolution parameter $\epsilon$, a value of rate-weakening ratio $b/a = 2$ and dimensionless spring stiffness $\kappa = 2 > \kappa_{crit}$. The panel (a) displays trajectories obtained with the slip law ($\epsilon \to 0$), while panel (d) with the aging law ($\epsilon = 1$). Panels (b) and (c), instead, display the elastic system response for two intermediate state evolution laws, respectively $\epsilon = 0.6$ and $0.7$. On each panel, the blue dashed line refers to steady state line, whereas the orange dashed lines refers to acceleration lines. Finally, the solid red line represents the critical trajectory that separates stable and unstable slip motion (see Equation \ref{eq: limiting trajectory2}), whereas the black solid lines are trajectories obtained numerically by solving the non-linear system of first-order differential equations (\ref{eq:psi_T}-\ref{eq:phi_T}) with initial conditions consistent with an initial velocity jump.}
   \label{fig: trajectories 3}
\end{figure}

When the condition (\ref{eq: critical trajectory condition}) is violated, slip motion is always stable. However, the spring-slider elastic system may undergo a transient of self-driven slip acceleration in response to perturbations. Indeed, following from (\ref{eq:phi_T}) and the definition of normalized state variable $\Theta$ 
\begin{equation}
\Theta = \frac{b}{a} \ln \left( \frac{V_* \theta}{D_c}\right)
\label{eq: normalized frictional state}
\end{equation}
for which the phase variables are related to it via $\Theta = \psi - \phi$, slip acceleration $\left(d \phi / d T >0\right)$ occurs when the state evolution $\dot{\Theta}$
\begin{equation}
\dot{\Theta} < \kappa \left( v_o - e^{\phi}\right)
\end{equation}
Since $v_o = 0$ in this case, the limit condition for slip acceleration reduces to 
\begin{equation}
\dot{\Theta}_a = - \kappa e^{\phi}
\end{equation}
which, with the help of (\ref{eq: intermediate state evolution law}), leads to the following condition for the dimensionless state variable $\Theta$
\begin{equation}
\Theta_a = -\frac{b}{a} \phi + \frac{b}{a} \ln \left[ \left( 1 - \frac{\epsilon \kappa}{b/a}\right)^{-1/\epsilon}\right]
\end{equation}
Now, since the dimensionless frictional stress $\psi$ and slip velocity jump $\phi$ are related to each others via $\psi = \phi + \Theta$, we can obtain a critical acceleration line in the phase plane $\left( \phi, \psi \right)$ that reads 
\begin{equation}
\psi_a = \left( 1 - \frac{b}{a}\right) \phi + \frac{b}{a} \ln \left[ \left( 1 - \frac{\epsilon \kappa}{b/a}\right)^{-1/\epsilon} \right]
\label{eq: acceleration line}
\end{equation}
The acceleration line in the phase plane - when it exists - is parallel to both the steady state line (\ref{eq: dimensionless steady state}) and the critical trajectory (\ref{eq: limiting trajectory2}). Moreover, its intersection with trajectories in the phase plane occurs at points on trajectories whose local tangent line is vertical (see for instance Figures \ref{fig: trajectories 2} and \ref{fig: trajectories 3}).\\
The necessary condition for transient acceleration is 
\begin{equation}
\frac{\epsilon \kappa}{b/a} < 1
\label{eq: acceleration condition}
\end{equation}
or, written dimensionally, when $k < b \sigma / (D_c \epsilon)$. When $\epsilon = 1$, we retrieve the results of \citet{HeSa09} for the aging law. Equation (\ref{eq: acceleration condition}) implies the existence of a minimum fault dimension (or perturbation wavelength) below which a transient of self-driven slip acceleration can never develop. Again, such a minimum dimension (or wavelength) decreases with decreasing values of $\epsilon$ parameter, vanishing in the limit of $\epsilon \to 0$. In the slip law limit, therefore, any perturbations that make the frictional stress exceed the value defined in (\ref{eq: acceleration line}), simply given by $\psi_a = \left( 1 - b/a\right)\phi + \kappa$ when $\epsilon \to 0$, induce a transient slip acceleration.\\

Now, similarly to section \ref{sec: non-stationary load point}, imagine that an external perturbation alters suddenly the motion of the slider, bringing instantaneously its slip velocity from the steady velocity $V_*$ to a new, larger value $V+\Delta V$. If the change in slip rate occurs faster than any appreciable state evolution can occur, then the dimensionless frictional stress $\psi$ increases by the same amount as that of the dimensionless slip velocity $\phi$. We can identify the critical velocity perturbation as $\Delta \phi_{crit} = \ln \left( 1 + \Delta V_{crit}/V_*\right)$ and determine its value as the distance between the origin (if we presume initial steady-state conditions) to the critical line $\psi_{crit} (\phi)$, given by (\ref{eq: limiting trajectory2}), along a $45^{\circ}$ incline.
Unlike the case of a non-stationary load point, here we are able to get an analytical expression for $\Delta \phi_{crit}$ that is valid for any $\epsilon \in \left[ 0,1\right]$ and reads
\begin{equation}
\Delta \phi_{crit} = \frac{1}{\epsilon} \cdot \ln \left( \frac{1}{1- \epsilon \cdot (\kappa/\kappa_{crit})}\right),
\label{eq: analytical phi crit}
\end{equation}
with its existence that still depends on the condition (\ref{eq: critical trajectory condition}) determining whether $\psi_{crit}(\phi)$ exists at all.

\begin{figure}[t!]
\centering
   \subfloat[]{
      \includegraphics[width=.48\textwidth]{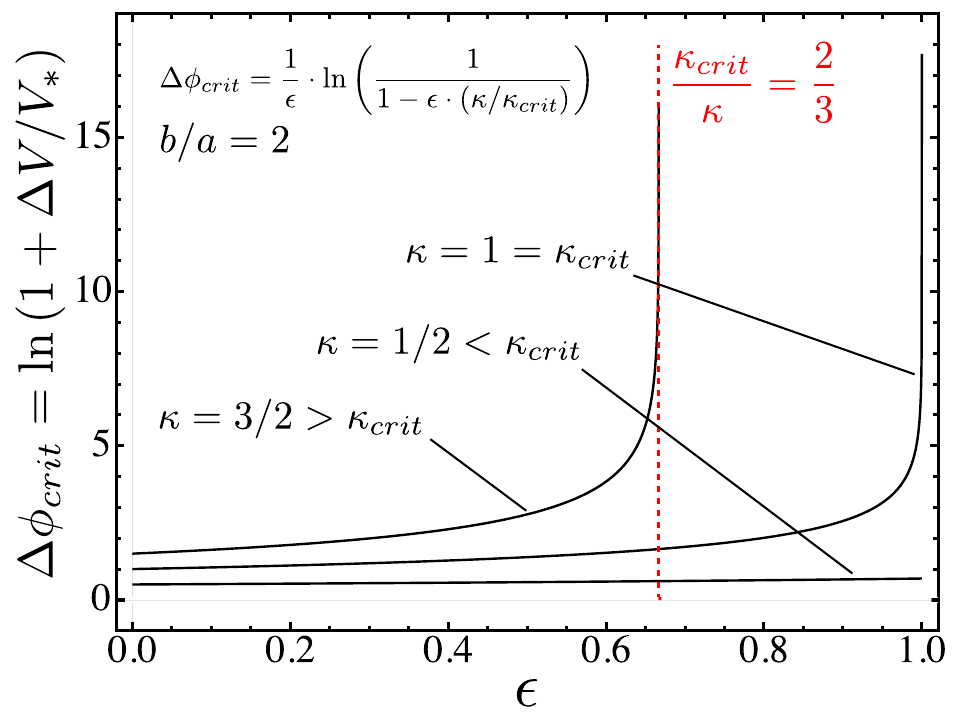}}
~
   \subfloat[]{
      \includegraphics[width=.48\textwidth]{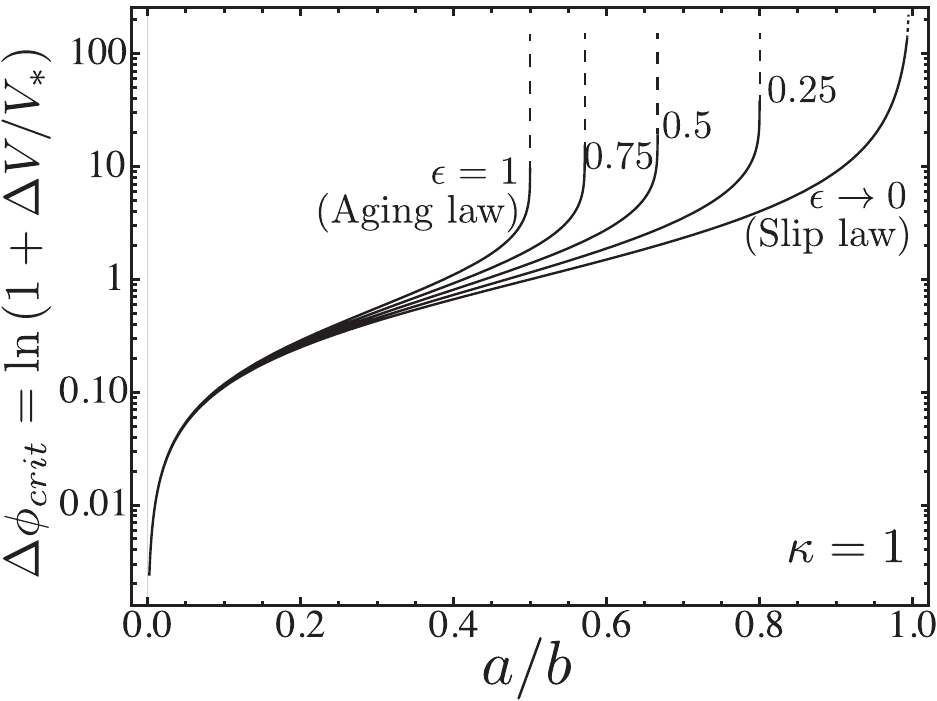}}
   \caption{For stationary loading ($v_o = 0$), Panel (a): Critical jump in sliding velocity leading to unstable sliding from steady-state initial conditions as a function of frictional state parameter $\epsilon$, for a given value of rate weakening ratio $b/a=2$. The black solid lines correspond to the analytical solution (\ref{eq: analytical phi crit}) for three values of spring stiffness $\kappa$:  below, equal and above the critical spring stiffness $\kappa_{crit}$.
 The vertical red dashed line refers to the critical value of $\epsilon$ parameter defined in Eq. (\ref{eq: critical epsilon}), and valid only when $\kappa > \kappa_{crit}$, i.e. when condition (\ref{eq: critical trajectory condition}) is satisfied. Panel (b): Same critical jump in sliding velocity from steady-state due to external perturbation but here as a function of the rate weakening ratio $a/b$, for a given value of stiffness $\kappa = 1$ and different fixed values of state evolution parameter $\epsilon$, ranging from aging ($\epsilon = 1$) to slip law ($\epsilon \to 0$) limits. For increasing values of $a/b$, the analytical curves asymptotically approach the critical value defined in (\ref{eq: critical ab}).}
\label{fig: limiting perturbation 2}
\end{figure}

When $\kappa < \kappa_{crit}$, a dynamic instability can be triggered for any value of frictional state evolution parameter $\epsilon$, with a critical jump in velocity that is always lower compared to any case when $\kappa \geq \kappa_{crit}$ (for the same frictional conditions). Furthermore, the minimum value of $\Delta \phi_{crit}$ is achieved in the slip law limit and is given by $\Delta \phi_{crit} = \kappa/\kappa_{crit}$, while the maximum value of $\Delta \phi_{crit}$ is reached in the aging law limit, for which $\Delta \phi_{crit} = \ln\left( \kappa_{crit} / (\kappa_{crit} -\kappa)\right)$. Figure \ref{fig: limiting perturbation 2}-a displays how $\Delta \phi_{crit}$ evolves as a function of $\epsilon$, for a given rate-weakening ratio $b/a = 2$ and three values of spring stiffness $\kappa$: lower, larger, and equal to the critical value $\kappa_{crit} = 1$. 
Examples of constant-$U$ trajectories in Figure \ref{fig: trajectories 2} obtained with $\kappa<\kappa_{crit}$ confirm the above considerations. A critical trajectory that divides stable and unstable slip motion can be clearly distinguished for any $\epsilon \in \left[ 0,1\right]$ (see red solid lines in Fig. \ref{fig: trajectories 2}  or in Fig. 2 of Suppl. Material where more results with intermediate state evolution laws are included). Moreover, the critical jump from the steady-state origin (denoted by the black point along the dashed blue line) slightly increases for increasing values of $\epsilon$, as highlighted by the length of the red arrows in panels (a-d).
Finally, we can also observe that the minimum ``kick" needed to induce slip acceleration is very similar for all the state evolution laws, although slightly larger finite-strength perturbations are needed in the aging law compared to the slip law. Both the critical trajectory $(U=0)$ and the acceleration line (orange dashed line) shift away from the steady state line as $\epsilon \to 1$; the region in the phase plane in which slip accelerates upon perturbation is wider for the aging law.

When the spring stiffness $\kappa = \kappa_{crit}$, the variation of the critical velocity jump $\Delta \phi_{crit}$ with $\epsilon$ parameter is very similar to the non-stationary load point scenario (see Fig. \ref{fig: limiting perturbation 1}): a singularity appears in the aging law limit. When $\epsilon = 1$, therefore, a dynamic instability can never be triggered regardless of the strength of the applied perturbation. Or, in other words, for slipping patches whose strength deviates from the aging law, there always exists a critical perturbation that can initiate a dynamic rupture regardless of whether the load point is stationary. The strength of such a perturbation decreases with decreasing values of $\epsilon$, reaching its minimum value in the slip law limit. The trajectories in the phase plane depicted in Figure 3 of Supplementary Material again illustrate these considerations.

Finally, when $\kappa > \kappa_{crit}$, Figure \ref{fig: limiting perturbation 2}-a shows that the singularity in the critical velocity jump $\Delta \phi_{crit}$ deviates from the aging law limit and is now centered at $\epsilon = \epsilon_{crit}$ defined by Eq. (\ref{eq: critical epsilon}). This implies that there exists a range of frictional state evolution laws $\left( \epsilon_{crit} < \epsilon \leq 1 \right)$ for which slip motion is unconditionally stable. The larger the spring stiffness is relative to the critical value $ \kappa_{crit}$, the wider is the stable interval. To illustrate this, we show in Figure \ref{fig: trajectories 3} how the trajectories in the phase plane change with different values of $\epsilon$, chosen to be below and above $\epsilon_{crit}$. We set $b/a= 2$ and $\kappa = 3/2 > \kappa_{crit}$, such that $\epsilon_{crit} = 2/3$. As an aside note also that with these input parameters, the necessary condition for transient slip acceleration along stable trajectories (\ref{eq: acceleration condition}) is satisfied for any $\epsilon \in \left[ 0,1\right]$.
From Figure $\ref{fig: trajectories 3}$ we can readily observe a change in elasto-frictional response when the $\epsilon$ parameter is below or above $\epsilon_{crit}$. 
In panels (a) and (b), which show trajectories associated with $\epsilon \to 0$ and $\epsilon  = 0.6$ respectively, there exists a critical line that divides stable and unstable slip motion (see the red solid line), with a jump from steady state that considerably increases as $\epsilon$ tends to $\epsilon_{crit}$ (see also Fig. 4 in Suppl. Material). The region in the phase plane in which stable slip accelerates transiently is larger when $\epsilon \to \epsilon_{crit}$ (from below), and the shift of the acceleration line from steady state increases with increasing value of $\epsilon$ (see Eq. (\ref{eq: acceleration line})). 
On the other hand, when $\epsilon > \epsilon_{crit}$ (see panels (c) and (d)), there are no unstable trajectories. All the orbits in the phase plane are characterized by $U<0$ and hence slip motion is always stable regardless of the magnitude of the jump from steady state. Large enough perturbations, therefore, can only induce transient slip acceleration in this case. Note, however, that if a stiffer spring is considered, the acceleration condition (\ref{eq: acceleration condition}) may not be satisfied for all $\epsilon > \epsilon_{crit}$. In Figure 5 of Supplementary Material, we present trajectories obtained with the same input parameters as above, but now the spring stiffness is larger and equal to $\kappa = 4 > \kappa_{crit} = 1$. We can observe that when $\epsilon > \epsilon_{crit} = 0.25$, the spring-slider motion is always stable and transient slip acceleration can only be induced when $\epsilon_{crit} < \epsilon < (b/a)/\kappa = 0.5$ (see panels (c-d)). 

So far, we have presented how the critical jump in slip velocity in response to perturbation varies with different state evolution laws and spring stiffnesses, while keeping constant the rate-weakening ratio $b/a$ (or its inverse). Frictional interfaces, however, exhibit different values of the parameter $a$ and $b$ depending on various factors, such as type of material of the two bodies in contact, temperature or values of effective normal stress on the interface (e.g., \cite{BlaMa98}). It is interesting to know, therefore, how $\Delta \phi_{crit}$ changes with the ratio $a/b$ for a particular state evolution law and spring stiffness. From Figure \ref{fig: limiting perturbation 2}-b, we observe that $\Delta \phi_{crit}$ decreases for more pronounced rate-weakening interfaces and that the state evolution law does not play a role when $a/b \to 0$. All the curves, indeed, collapsed into one curve that asymptotically goes to 0 when $a/b$ approaches to 0. An instability can thus be triggered with small perturbations in slip motion in that limit.
On the contrary, for increasing values of $a/b$, $\Delta \phi_{crit}$ increases for any $\epsilon \in \left[ 0,1\right]$. However, a particular state evolution law defines the range of $a/b$ in which such evolution occurs, with the maximum asymptotic value given by (\ref{eq: critical ab}) (for a given $\epsilon$ and spring stiffness $\kappa$). With the aging law, therefore, there exists an interval of $a/b$ for which slip motion is unconditionally stable. This interval, however, vanishes in the slip law limit: a sufficiently large perturbation can trigger an instability for any value of $a/b \in \left[0, 1\right)$. Note that $\Delta \phi_{crit} \to \infty$ when $a/b\to 1$, i.e. on rate-neutral interfaces an instability can never develop regardless of the strength of the perturbation applied.

\section{Equivalence between the intermediate and Nagata state evolution laws}
Following a series of ``stress-step" experiments using a double direct shear apparatus with pre-cut granite samples, \citet{NaNa12} proposed a revised version of rate-and-state friction law that includes a dependence on stressing rate in the form
\begin{align}
\begin{split}
&f = f_* + a_N \ln \left( \frac{V}{V_*}\right) + b_{N} \ln \left( \frac{V_* \theta}{D_{c_N}}\right)\\
& \frac{\partial \theta}{\partial t} = 1 - \frac{V \theta}{D_{c_N}} - \frac{c_N \theta}{\sigma b_N} \frac{\partial \tau}{\partial t}
\label{eq: Nagata law}
\end{split}
\end{align}
where the subscript $_{N}$ distinguishes the Nagata parameters from those of classical rate-and-state friction law presented in sub-section \ref{subsec: RateAndState}. 
If $c = 0$, the Nagata law reduces to the aging law (\ref{eq: aging law}). 

The Nagata evolution law is equivalent to the intermediate law (\ref{eq: intermediate state evolution law}). Specifically, the parameters of the intermediate law $\left( D_c, a, b, \epsilon \right)$ are uniquely determined by the choice of the Nagata law parameters $\left( D_{c_N}, a_N, b_N, c_N\right)$: 
 \begin{align}
\begin{split}
&D_c = \frac{D_{c_N}}{1 + c_N}\\
& a = \frac{a_N}{1+c_N}\\
& \epsilon = 1 - \frac{a_N c_N}{(1+c_N) b_N}\\
& b = \left( 1 - \frac{a_N c_N}{(1+c_N) b_N}\right) b_N
\label{eq: mapping Nagata and intermediate evolution law}
\end{split}
\end{align}
This equivalency was independently recognized by Hiroyuki Noda (personal communication, 2018) and further discussed by \citet{NoCha23}. 

\section{Conclusion}
\label{sec:conclusions}
We have studied non-linear slip motion stability analysis of a single-degree-of-freedom elastic system with different frictional state evolution laws. Using the evolution law (\ref{eq: intermediate state evolution law}) and its parametrization via a dimensionless parameter $\epsilon$, originally proposed by \citet{Ruin83}, we have investigated analytically and numerically the elastic response of the spring-slider system in the parameter space identified by scaling analysis. Our results extend and generalize those of \citet{GuRi84} and \citet{RaRi99}, valid respectively only for the slip and aging laws, and provide a complete picture of slip motion stability in the non-linear regime for a range of realistic frictional state evolutions. 

We have demonstrated that the critical stiffness (and by analogy to continuum faults, the nucleation length) for slip instability is not well-defined for a wide range of friction descriptions and quantified the conditions that initiate an instability as function of the governing dimensionless parameters. Remarkably, our results show that slipping patches whose state evolve following the slip law are more susceptible to transient states of self-driven slip accelerations or dynamic instabilities than those whose state evolves in an aging-law-like fashion. Moreover, a dynamic instability can nucleate on smaller faults, provided that sufficiently large perturbations are applied to the system and frictional state deviates, even slightly, from the aging law.
This loading rate dependency on the nucleation process agrees with experimental observations on spatially extended frictional interfaces in which reduced dynamic-rupture nucleation lengths have been attributed to sudden, localized increases in loading rate associated with local strength heterogeneities along the frictional interface or with anthropogenic fluid injection \citep{McLaYama17, XuFu18, McLaskey19, GueNie19, GoRu21, JiWa22}.
Despite the model simplicity and the simplifying assumptions, the analytical results presented here may provide support to observations of laboratory earthquake experiments and numerical findings obtained with more complex continuum models.

\section*{CrediT authorship contribution statement}
\textbf{Federico Ciardo}: Formal analysis, Investigation, Software, Methodology, Visualization, Writing - original draft, Writing - review \& editing.\\
\textbf{Robert C. Viesca}: Formal analysis, Conceptualization, Funding acquisition, Writing - review \& editing, Supervision.

\section*{Declaration of competing interest}
The authors declare that they have no known competing financial interests or personal relationships that could have appeared to influence the work reported in this paper. 

\section*{Data availability}
No data were used for the research described in this article. 

\section*{Acknowledgments}
This work was supported by the National Science Foundation (grant EAR-1834696).

\begin{appendices}
\section{Appendix: Elasto-frictional response of spring-slider system subjected to finite perturbations}
\label{app: appendix}

\subsection*{$\kappa = \kappa_{crit}$: periodic response}

Results from linear stability analysis, valid for both slip and aging law, showed that at neutral stability the elastic response remains quasi-static and periodic upon small perturbations in slip motion \citep{Ruin83, RiRu83}. Here, we show that this is also valid in the non-linear regime with an intermediate state evolution law, provided that the applied perturbation is not large enough to trigger a dynamic instability (as discussed in sub-section \ref{subsec: k = k critical}). 

We consider the single-degree-of-freedom spring-slider elastic system depicted in Figure (\ref{fig: spring-slider}) that slides on a rate weakening interface characterized by $b/a = 2$, with frictional state evolution that follows an intermediate law ($\epsilon = 0.25$). We assume that the spring stiffness $\kappa$ is equal to the critical value $\kappa_{crit} = b/a-1 = 1$ and consider small perturbation in the form of a sudden increase of load point rate from steady-state value. Specifically, we assume a velocity jump perturbation of $\Delta v_o / v_o = 1.33$ that is slightly below the critical value $\left(\Delta v_o / v_o\right)_{crit} = 1.34$ needed to initiate a dynamic instability (see Figure \ref{fig: stability diagram}-a). Using as initial conditions the steady state conditions prior the velocity jump, i.e. $\psi = 0$ and $\phi=0$, we integrate numerically Equations (\ref{eq:psi_T}-\ref{eq:phi_T}) for the evolution of normalized shear stress $\psi$ and log of slip velocity $\phi$ (relative to steady state), in function of slip displacement $\delta/D_c$ since velocity jump. 

\begin{figure}[t!]
\centering
   \subfloat[]{
      \includegraphics[width=.35\textwidth]{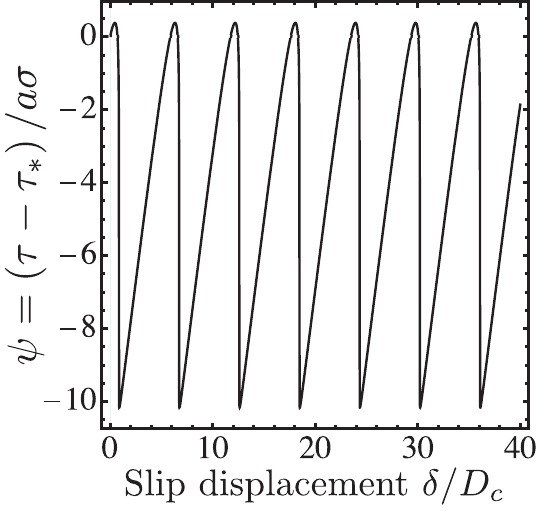}}
~
   \subfloat[]{
      \includegraphics[width=.35\textwidth]{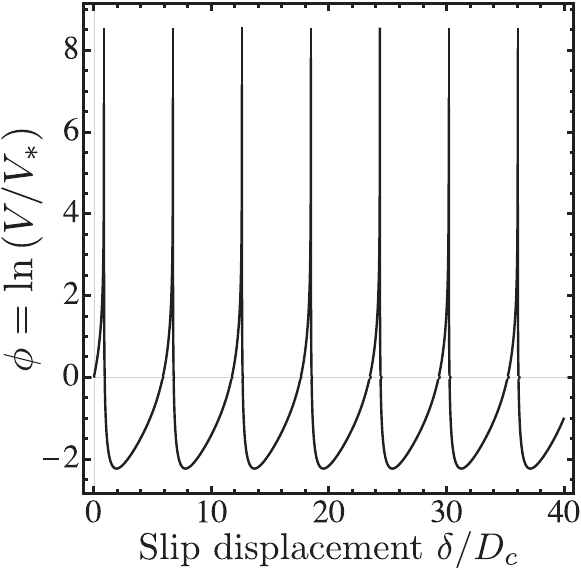}}
\\
   \subfloat[]{
      \includegraphics[width=.35\textwidth]{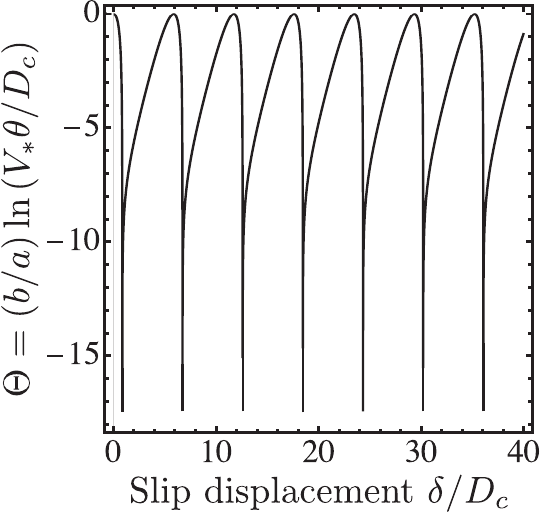}}
~
   \subfloat[]{
      \includegraphics[width=.35\textwidth]{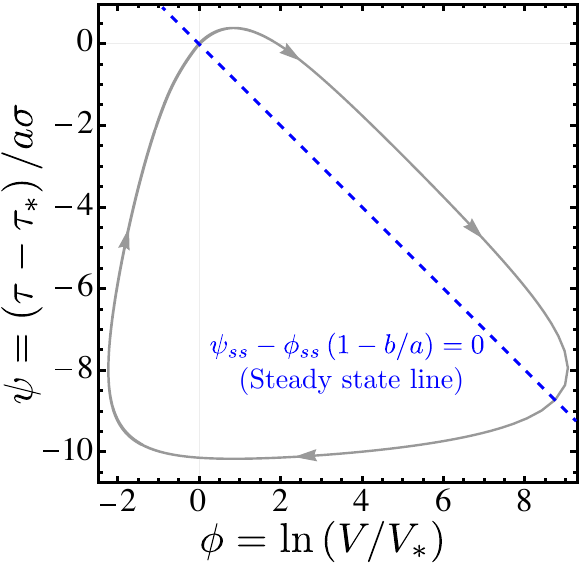}}
   \caption{Example of stable and periodic elastic response of the spring-slider when the spring stiffness $\kappa$ is equal to the critical value $\kappa_{crit}$ and a small load point velocity perturbation is applied to the system. In this particular example, the frictional state follows an intermediate evolution law characterized by $\epsilon = 0.25$, the rate weakening ratio is $b/a = 2$ and hence $\kappa = \kappa_{crit} = 1$, and the load point velocity perturbation is $\Delta v_o / v_o = 1.33$ that is slightly below the critical value $\left(\Delta v_o / v_o\right)_{crit} = 1.34$. Panels (a-c) display the evolution of normalized shear stress $\psi$, slip velocity $\phi$ and frictional state $\Theta$ as function of normalized slip accumulation $\delta/D_c$ since velocity jump. Panel (d), instead, displays the corresponding elastic response in the phase plane.}
   \label{fig: periodic motion}
\end{figure}

Numerical results are displayed in Figure \ref{fig: periodic motion}. Panels (a), (b) and (c) display clear periodic oscillations of shear stress $\psi$, slip velocity $\phi$ and normalized frictional state $\Theta = (b/a) \ln \left(\theta V_*/D_c\right)$ with slip accumulation respectively. The corresponding trajectory in the phase plane $\left( \phi, \psi \right)$ forms a closed contour (see panel (d)). 

\begin{figure}[t!]
\centering
\includegraphics[width=.9\textwidth]{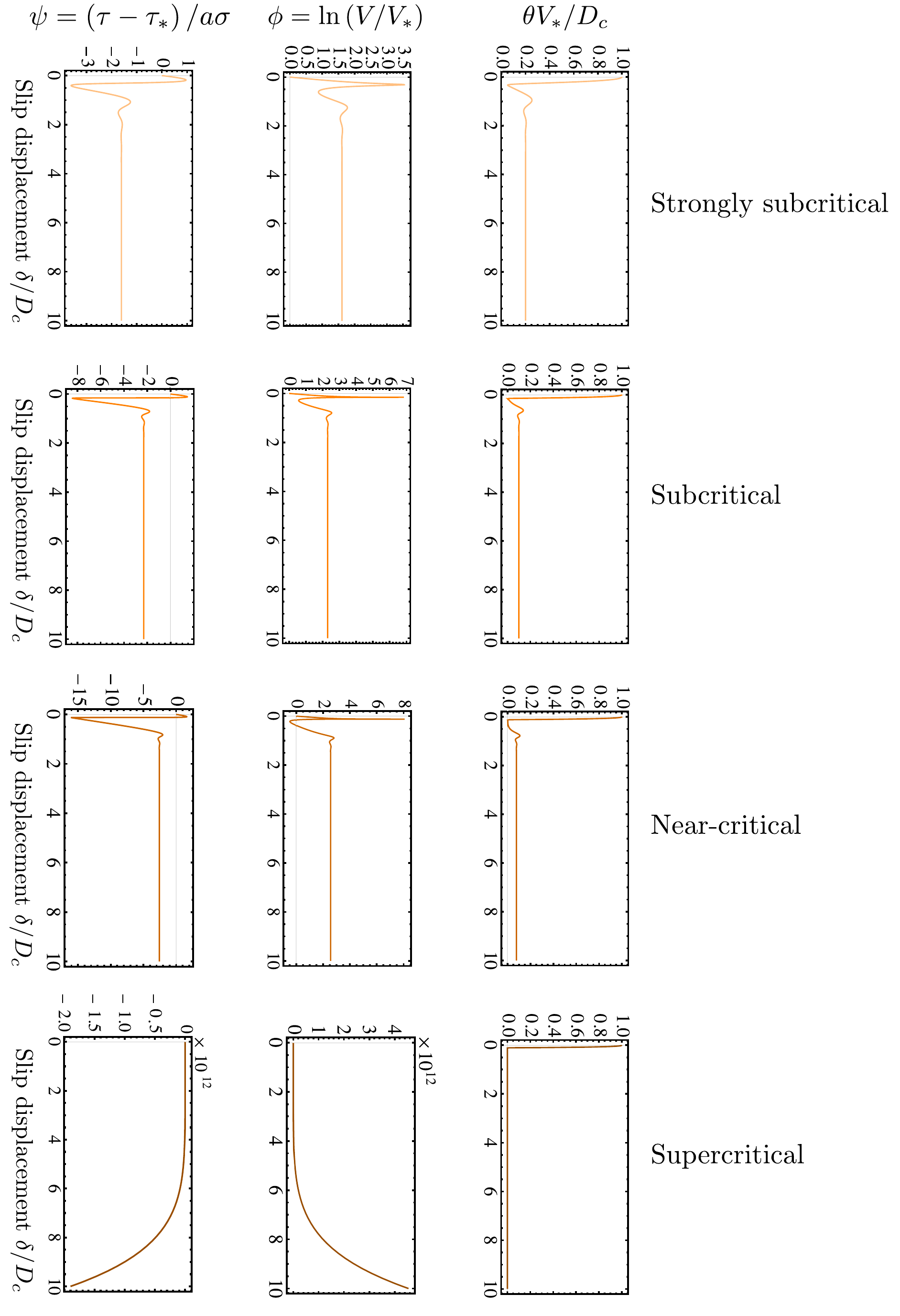}
\caption{From left to right: evolution of normalized shear stress $\psi$, slip velocity $\phi$ and frictional state $e^{\Theta /(b/a)} = \theta V_*/D_c$ as function of slip accumulation $\delta/D_c$ when subcritical, near-critical and supercritical load point velocity perturbations are applied to the spring-slider elastic system with stiffness $\kappa$ larger than $\kappa_{crit}$. Each row corresponds to the specific scenario reported in Figure \ref{fig: stability diagram}-b - note also that the same color has been used.}
\label{fig: non-periodic motion}
\end{figure}

\subsection*{$\kappa \neq \kappa_{crit}$:  non-periodic response}
Finally, we present similar responses of spring-slider system subjected to finite perturbations in point loading rate, but now the spring stiffness $\kappa$ is not equal to the critical value $\kappa_{crit}$. As discussed in subsection \ref{subsec: k not = k critical}, in this condition the elastic response is never periodic, but can be stable (with decaying oscillations amplitude) or unstable (with diverging slip rate) depending on the strength of the applied perturbation (see stability diagram in Figure \ref{fig: stability diagram}-a). Referring to the specific scenario reported in Figure \ref{fig: stability diagram}-b, which shows the elastic response in the phase plane when subcritical, near-critical and supercritical perturbations are applied to the system, we present in Figure \ref{fig: non-periodic motion} the corresponding evolution of $\psi$, $\phi$ and $e^{\Theta/(b/a)}= \theta V_*/D_c$ as function of slip displacement $\delta/D_c$. Consistently with the trajectories of Fig. \ref{fig: stability diagram}-b, we can observe that shear stress, the slip velocity as well as the frictional state evolve toward new steady state values after sufficiently slip displacement, when subcritical or near-critical velocity jumps are applied to the elastic system. Conversely, a supercritical perturbation leads to a rapid and continuous loss of frictional strength and - as a result - a diverging slip rate. A new steady state condition is thus never achieved.  

\section{Appendix: Equivalent stiffness in a continuum deformable interface}
\label{app: appendix 2}
 
 Here we provide details on how an effective stiffness $k/k_*$ can be determined for a continuum frictional interface via the solution of an associated eigenvalue problem. The eigenvalues and corresponding eigenmodes are determined by the geometry of the interface and the manner of loading. The eigenvalues correspond to the effective stiffness of each eigenmode, and may take on a discrete or continuous range of values, according to the geometry considered. Generally, when the interface is unbounded, these stiffnesses will be continuous; when the interface is finite, these will be a discrete set. 

\paragraph{Case (A): Infinite interface between two unbounded elastic half-spaces} Let us consider an infinitely-long planar interface that divides two linear-elastic half-spaces characterized by shear modulus $\mu$ and Poisson's ratio $\nu$ (see sketch in Figure (\ref{fig: continuum models})). The linear operator that determines the elastic change in shear stress in response to a given slip distribution $\delta (x)$ in this case is   
\begin{equation}
 \mathcal{L}\left( \delta\right) = \dfrac{\mu^{\prime}}{\pi} \int_{-\infty}^{\infty} \dfrac{d \delta/d s}{x-s} \, \text{ds}
 \label{eq: operator app 1}
\end{equation}
where $\mu^{\prime}$ is an equivalent shear modulus that depends on the mode of deformation: $\mu^{\prime} = \mu/2 (1-\nu)$ in mode II (in-plane) and $\mu^{\prime}=\mu/2$ in mode III (anti-plane). 
Using shorthand notation, we can write this as
\begin{equation}
\mathcal{L}(\delta) = \mu^{\prime} \mathcal{H}(d \delta/d s)
\end{equation}
where $\mathcal{H}(\cdot)$ is the Hilbert transform convolution for the operator (\ref{eq: operator app 1}). We may write an eigenequation in the form of (\ref{eq: eigenequation}) as
\begin{equation}
\mu^{\prime} \mathcal{H}(d \delta/d s) = k \delta,
\label{eq: eq1}
\end{equation}
where $k$ is an eigenvalue (eigenstiffness) that has units of stress per unit distance.\\ 
Given that the Hilbert transform has the property that $\mathcal{H} \left( \mathcal{H}\left( f\right)\right) = -f$, we may rewrite (\ref{eq: eq1}) as
\begin{equation}
- \mu^{\prime} \frac{d \delta}{d x} = k \mathcal{H}\left(\delta\right)
\label{eq: unknown 1}
\end{equation}
In taking the derivative of (\ref{eq: unknown 1}) and using (\ref{eq: eq1}), we can re-express the eigenvalue problem as 
\begin{equation}
\frac{d^2 \delta}{d x^2}+ \left( \frac{k}{\mu^{\prime}}\right)^2 \delta = 0,
\label{eq: final equation 1}
\end{equation}
which has solutions $\delta (x) = e^{i (k/\mu^{\prime}) x}$. Relating the wavenumber $k/\mu^{\prime}$ to a wavelength $\lambda$ as $\dfrac{k}{\mu^{\prime}} = \dfrac{2 \pi}{\lambda}$, we may alternatively express the eigenstiffness as 
\begin{equation}
k = 2 \pi \cdot \frac{\mu^\prime}{\lambda},
\end{equation}
which continuously varies with the choice of $\lambda$.

\paragraph{Case (B): Finite-length interface in an unbounded elastic medium} Let us consider now a similar scenario to the previous one, with the only difference being that the sliding interface is now finite in one direction and characterized by a half-length $L$ (see Figure \ref{fig: continuum models}, case (B)). The corresponding elastic operator now is not defined over the infinite line, but it operates over the finite interface via the following boundary integral 
\begin{equation}
 \mathcal{L}\left( \delta\right) = \dfrac{\mu^{\prime}}{\pi} \int_{-L}^{L} \dfrac{d \delta/d s}{x-s} \,\text{ds}
\end{equation} 
Upon scaling the spatial coordinates by $L$ and rearranging a few terms, the eigenequation (\ref{eq: eigenequation}) reads
\begin{equation}
\frac{k}{(\mu^{\prime}/L)} \delta =  \frac{1}{\pi} \int_{-1}^{1} \dfrac{d \delta/d s}{x-s} \,\text{d}\text{s}
\label{eq: eigenvalue problem Uenishi-Rice}
\end{equation}
where spatial coordinates $x, s$ are implicitly scaled by $L$. The eigenvalue problem (\ref{eq: eigenvalue problem Uenishi-Rice}) is exactly that resulting from an instability analysis on a slip-weakening fault under plane-strain condition, and whose solution has already been obtained numerically by \citet{UeRi03}. Here, we solve this eigenproblem semi-analytically using a Gauss-Chebishev quadrature with 500 integration points (see e.g. \citep{ViGa18}). The calculated smallest eigenvalue is
\begin{equation}
k = \left(1.157773882\dots\right) \cdot \frac{\mu^\prime}{L},
\end{equation}
with the associated eigenmode that is plotted in Figure \ref{fig: continuum models}, case (B) (see red curve).

\paragraph{Case (C): Infinite interface near and parallel to a free-surface of an unbounded elastic half-space} Unlike the last two scenarios, here we consider an infinitely-long planar interface that is parallel to a free-surface of an unbounded linear-elastic half-space (see Figure \ref{fig: continuum models}, case (C)). We further assume that its offset from the free-surface $h$ is much smaller than the lengthscales associated with variations in distance, such that this model reduces to that of a one-dimensional compressible slab sliding on a rigid base. The elastic operator in this case reads (see supplementary materials of \citet{Vies16a})
\begin{equation}
\mathcal{L}(\delta) = - E^{\prime} h \frac{d^2 \delta}{d x^2},
\label{eq: fin eq}
\end{equation}
where $E^{\prime}$ is an equivalent modulus that depends on deformation mode ($E^{\prime} = 2\mu / (1-\nu)$ in mode II and $E^{\prime} = \mu$ in mode III) and $x$ spans the interval $\left(-\infty; +\infty\right)$. In this particular case, the eigenequation (\ref{eq: eigenequation}) reduces to
\begin{equation}
\frac{d^2 \delta}{d x^2} + \frac{k}{E^\prime h} \delta = 0,
\label{eq: final equation 2}
\end{equation}
which has solutions $\delta (x)  = e^{i \sqrt{\frac{k}{E^\prime h}}x}$, and again, relating the wavenumber of this solution to the wavelength $\lambda$ via $\sqrt{\dfrac{k}{E^\prime h}}=\dfrac{2 \pi}{\lambda}$, we may rewrite the eigenstiffness as 
\begin{equation}
k = 4 \pi^2 \cdot \frac{E^{\prime} h }{\lambda^2}
\end{equation}

\paragraph{Case (D): Finite-length interface near and parallel to a free-surface of an unbounded elastic half-space} 
Perturbing the previous scenario, let the frictional interface be finite and characterized by a half-length $L$. The elastic operator $\mathcal{L}(\delta)$ remains (\ref{eq: fin eq}), as well as the eigenequation (\ref{eq: final equation 2}), except that now the slipping coordinate spans a finite interval $\left[ -L; +L\right]$. Here, however, unlike to the infinite interface case, the finiteness of the slipping region requires boundary conditions for $\delta$. An appropriate choice of boundary conditions is that eigenmodes vanish at the endpoints: $\delta \left( \pm L\right) = 0$. Considering that solutions to (\ref{eq: final equation 2}) $e^{i \sqrt{\frac{k}{E^\prime h}}x}$ can be decomposed into eigenmodes that are symmetric or antisymmetric about the origin, $\cos\left( \sqrt{\dfrac{k}{E^\prime h}} x\right)$ or $\sin\left( \sqrt{\dfrac{k}{E^\prime h}} x\right)$, imposing the boundary conditions amounts to requiring that $\sqrt{\dfrac{k}{E^\prime h}} L = n \dfrac{\pi}{2}$, with $n$ being an integer. The smallest eigenstiffness is hence
\begin{equation}
k = \frac{\pi^2}{4} \cdot  \frac{E^\prime h}{L^2}
\end{equation}

\paragraph{Case (E): Circular interface in an unbounded elastic medium} Finally, we present the case of a circular interface with radius $R$ embedded in a three-dimensional linear-elastic unbounded domain. The operator in this particular scenario reads
\begin{equation}
\mathcal{L}\left( \delta\right) = \dfrac{\mu}{2 \pi} \int_{0}^{R} \dfrac{\partial \delta}{\partial s} \left( \dfrac{E\left[ f(r/s)\right]}{r-s} - \dfrac{F\left[ f(r/s)\right]}{s+r}\right)  \text{ds},
\label{eq: functional circular}
\end{equation}
where $E$ and $F$ are the complete elliptic integrals of second and first kind respectively, and $f(U) = 2 \sqrt{U}/(1+U)$. This representation is valid for a Poisson ratio $\nu=0$, but can be considered as a leading-order approximation for a case where $\nu\neq0$.
Upon mapping the above integral from $r, s \in \left[ 0; R\right]$ to $\bar{r}, \bar{s} \in \left[ -1; 1\right]$ using the following coordinate transformations
\begin{equation}
s = \frac{R}{2} \left( \bar{s} + 1\right) \quad r = \frac{R}{2} \left( \bar{r}+1\right),
\end{equation}
and scaling slip with a characteristic scale $\delta_*$, the eigenequation (\ref{eq: eigenequation}) reduces to 
\begin{equation}
\frac{k}{\mu/R} \cdot \bar{\delta} = \frac{1}{\pi}  \int_{-1}^{1} \dfrac{\partial \bar{\delta}}{\partial \bar{s}} \left( \dfrac{1}{\bar{r}-\bar{s}} E\left[ f\left(\frac{\bar{r}+1}{\bar{s}+1}\right)\right] - \dfrac{1}{\bar{r}+\bar{s}+2} F\left[ f\left(\frac{\bar{r}+1}{\bar{s+1}}\right)\right] \right)  \text{d}\bar{\text{s}},
\label{eq: eigenequation circular}
\end{equation}
which can be readily solved for the eigenvalues $\dfrac{k}{\mu/R}$ and the associated eigenmodes $\bar{\delta} = \delta / \delta_*$. We use Gauss-Chebyshev quadrature for the above integral representation, considering a discretization on a primary and a complimentary sets of nodes that correspond to the roots of the Chebyshev polynomials of third and fourth kind respectively, and solve the eigenvalue problem semi-analytically \citep{ViGa18}. It is worth noting that the elastic kernel in (\ref{eq: eigenequation circular}) or (\ref{eq: functional circular}) is to the leading order of Cauchy type. However, we acknowledge the presence of a weaker log singularity that compromise the accuracy of the numerical solution if not appropriately taken care of. To improve the accuracy of the quadrature, we follow a similar approach proposed by \citet{LiLe19} to regularize the logarithmic singularity.
Using 500 Gauss-Chebyshev points, the smallest eigenvalue in this case is
\begin{equation}
k = \left(1.003059516\dots \right) \cdot \frac{\mu}{R}
\end{equation}

\end{appendices}


\newpage 

\begin{thebibliography}{}

\bibitem[Ampuero and Rubin, 2008]{AmRu08}
Ampuero, J.-P. and Rubin, A.~M. (2008).
\newblock Earthquake nucleation on rate and state faults -- aging and slip
  laws.
\newblock {\em Journal of Geophysical Research: Solid Earth}, 113(B1).
\newblock \url{https://doi.org/10.1029/2007JB005082}.

\bibitem[Bhattacharya et~al., 2017]{BhaRu17}
Bhattacharya, P., Rubin, A.~M., and Beeler, N.~M. (2017).
\newblock Does fault strengthening in laboratory friction experiments really
  depend primarily upon time and not slip?
\newblock {\em Journal of Geophysical Research}, 122:6389--6430.
\newblock \url{https://doi.org/10.1002/2017JB013936}.

\bibitem[Bhattacharya et~al., 2022]{BhaRu22}
Bhattacharya, P., Rubin, A.~M., Tullis, T.~E., Beeler, N.~M., and Okazaki, K.
  (2022).
\newblock The evolution of rock friction is more sensitive to slip that elasped
  time, even at near-zero slip rates.
\newblock {\em Earth, Atmospheric, and Planetary Science}, 119(30).
\newblock \url{https://doi.org/10.1073/pnas.2119462119}.

\bibitem[Blanpied et~al., 1998]{BlaMa98}
Blanpied, M.~L., Marone, C.~J., Lockner, D.~A., Byerlee, J.~D., and King, D.~P.
  (1998).
\newblock Quantitative measure of the variation in fault rheology due to
  fluid-rock interactions.
\newblock {\em Journal of Geophysical Research}, 103(B5):9691--9712.
\newblock \url{https://doi.org/10.1029/98JB00162}.

\bibitem[Byerlee, 1970]{Byerlee1970}
Byerlee, J.~D. (1970).
\newblock The mechanics of stick-slip.
\newblock {\em Tectonophysics}, 9:475--486.
\newblock \url{https://doi.org/10.1016/0040-1951(70)90059-4}.

\bibitem[Cattania, 2023]{Cattania23}
Cattania, C. (2023).
\newblock A source model for earthquakes near nucleation dimension.
\newblock {\em Bull. Seism. Soc. Am.}, 113(3).
\newblock \url{https://doi.org/10.1785/0120220045}.

\bibitem[Dieterich, 1978]{Diet78}
Dieterich, J.~H. (1978).
\newblock Time-dependent friction and the mechanics of stick-slip.
\newblock {\em Pure and Applied Geophysics}, 116(4-5):790--806.
\newblock \url{https://doi.org/10.1007/BF00876539}.

\bibitem[Dieterich, 1979]{Diet79}
Dieterich, J.~H. (1979).
\newblock Modeling of rock friction: 1. experimental results and constitutive
  equations.
\newblock {\em Journal of Geophysical Research: Solid Earth},
  84(B5):2161--2168.
\newblock \url{https://doi.org/10.1029/JB084iB05p02161}.

\bibitem[Dieterich, 1992]{Diet92}
Dieterich, J.~H. (1992).
\newblock Earthquake nucleation on faults with rate-and-state-dependent
  strength.
\newblock {\em Tectonophysics}, 211:115--134.
\newblock \url{https://doi.org/10.1016/0040-1951(92)90055-B}.

\bibitem[Garagash, 2021]{Ga21}
Garagash, D.~I. (2021).
\newblock Fracture mechanics of rate-and-state faults and fluid injection
  induced slip.
\newblock {\em Phil. Trans. R. Soc. A}, 379.
\newblock \url{https://doi.org/10.1098/rsta.2020.0129}.

\bibitem[Gori et~al., 2021]{GoRu21}
Gori, M., Rubino, V., Rosakis, A.~J., and Lapusta, N. (2021).
\newblock Dynamic rupture initiation and propagation in a fluid-injection
  laboratory setup with diagnostics across multiple temporal scales.
\newblock {\em Earth, Atmospheric, and Planetary Science}, 118(51).
\newblock \url{https://doi.org/10.1073/pnas.2023433118}.

\bibitem[Gu et~al., 1984]{GuRi84}
Gu, J.-C., Rice, J.~R., Ruina, A.~L., and Tse, S.~T. (1984).
\newblock Slip motion and stability of a single degree of freedom elastic
  system with rate and state dependent friction.
\newblock {\em Journal of the Mechanics and Physics of Solids}, 32(3):167--196.
\newblock \url{https://doi.org/10.1016/0022-5096(84)90007-3}.

\bibitem[Gu and Wong, 1991]{GuWo91}
Gu, Y. and Wong, T.-F. (1991).
\newblock Effects of loading velocity, stiffness, and inertia on the dynamics
  of a single degree of freedom spring-slider system.
\newblock {\em Journal of Geophysical Research: Solid Earth},
  96(B13):21677--21691.
\newblock \url{https://doi.org/10.1029/91JB02271}.

\bibitem[Gu{\'e}rin-Marthe et~al., 2019]{GueNie19}
Gu{\'e}rin-Marthe, S., Nielsen, S., Bird, R., Giani, S., and Di~Toro, G.
  (2019).
\newblock Earthquake nucleation size: evidence of loading rate dependence in
  laboratory faults.
\newblock {\em Journal of Geophysical Research}, 124(689-708).
\newblock \url{https://doi.org/10.1029/2018JB016803}.

\bibitem[Gvirtzman and Fineberg, 2021]{GviFin21}
Gvirtzman, S. and Fineberg, J. (2021).
\newblock Nucleation fronts ignite the interface rupture that initiates
  frictional motion.
\newblock {\em Nature Physics}, 17:1037--1042.
\newblock \url{https://doi.org/10.1038/s41567-021-01299-9}.

\bibitem[Helmstetter and Shaw, 2009]{HeSa09}
Helmstetter, A. and Shaw, B.~E. (2009).
\newblock Afterslip and aftershocks in the rate-and-state friction law.
\newblock {\em Journal of Geophysical Research}, 114(B01308).
\newblock \url{https://doi.org/10.1029/2007JB005077}.

\bibitem[Ji et~al., 2022]{JiWa22}
Ji, Y., Wang, L., Hofmann, H., Kwiatek, G., and Dresen, G. (2022).
\newblock High-rate fluid injection reduces the nucleation length of laboratory
  earthquakes on critically stressed faults in granite.
\newblock {\em Geophys. Res. Letters}, 49.
\newblock \url{https://doi.org/10.1029/2022GL100418}.

\bibitem[Kaneko et~al., 2016]{KaNie16}
Kaneko, Y., Nielsen, S., and Carpenter, B.~M. (2016).
\newblock The onset of laboratory earthquakes explained by nucleating rupture
  on a rate-and-state fault.
\newblock {\em Journal of Geophysical Research}, 121(8):6071--6091.
\newblock \url{https://doi.org/10.1002/2016JB013143}.

\bibitem[Latour et~al., 2013]{LaScu13}
Latour, S., Schubnel, A., Nielsen, S., Madariaga, R., and Vinciguerra, S.
  (2013).
\newblock Characterization of nucleation during laboratory earthquakes.
\newblock {\em Geophys. Res. Letters}, 40:5064--5069.
\newblock \url{https://doi.org/10.1002/grl.50974}.

\bibitem[Liu et~al., 2019]{LiLe19}
Liu, D., Lecampion, B., and Garagash, D.~I. (2019).
\newblock Propagation of a fluid-driven fracture with fracture-length dependent
  apparent toughness.
\newblock {\em Engineering Fracture Mechanics}, 220:106616.
\newblock \url{https://doi.org/10.1016/j.engfracmech.2019.106616}.

\bibitem[Marone, 1998]{Ma98}
Marone, C. (1998).
\newblock Laboratory-derived friction laws and their application to seismic
  faulting.
\newblock {\em Annu. Rev. Earth Planet. Sci.}, 26:643--696.
\newblock \url{https://doi.org/10.1146/annurev.earth.26.1.643}.

\bibitem[Marty et~al., 2023]{MaSch2023}
Marty, S., Schubnel, A., Bhat, H.~S., Fukuyama, E., Latour, S., Nielsen, S.,
  and Madariaga, R. (2023).
\newblock Nucleation of laboratory earthquakes: quantitative analysis and
  scalings.
\newblock {\em Journal of Geophysical Research}, 128(3).
\newblock \url{https://doi.org/10.1029/2022JB026294}.

\bibitem[McLaskey, 2019]{McLaskey19}
McLaskey, G.~C. (2019).
\newblock Earthquake initiation from laboratory observations and implicaitons
  for foreshocks.
\newblock {\em Journal of Geophysical Research: Solid Earth},
  124(12):12882--12904.
\newblock \url{https://doi.org/10.1029/2019JB018363}.

\bibitem[McLaskey and Yamashita, 2017]{McLaYama17}
McLaskey, G.~C. and Yamashita, F. (2017).
\newblock Slow and fast ruptures on a laboratory fault controlled by loading
  characteristics.
\newblock {\em Journal of Geophysical Research: Solid Earth}, 122:3719--3738.
\newblock \url{https://doi.org/10.1002/2016JB013681}.

\bibitem[Nagata et~al., 2012]{NaNa12}
Nagata, K., Nakatani, M., and Yoshida, S. (2012).
\newblock A revised rate- and state-dependent friction law obtained by
  constraining constitutive and evolution laws separately with laboratory data.
\newblock {\em Journal of Geophysical Research}, 117(B02314).
\newblock \url{https://doi.org/10.1029/2011JB008818}.

\bibitem[Nielsen et~al., 2010]{NiTa10}
Nielsen, S., Taddeucci, J., and Vinciguerra, S. (2010).
\newblock Experimental observations of stick-slip instability fronts.
\newblock {\em Geophys. J. Int.}, 180:697--702.
\newblock \url{https://doi.org/10.1111/j.1365-246X.2009.04444.x}.

\bibitem[Noda and Chang, 2023]{NoCha23}
Noda, H. and Chang, C. (2023).
\newblock Tertiary creep behvior for various rate- and state-dependent friction
  laws.
\newblock {\em Earth and Planetary Science Letters}, 619(118314).
\newblock \url{https://doi.org/10.1016/j.epsl.2023.118314}.

\bibitem[Ohnaka and Shen, 1999]{OhnaShe99}
Ohnaka, M. and Shen, L.-F. (1999).
\newblock Scaling of the shear rupture process from nucleation to dynamic
  propagation: {I}mplications of geometric irregularity of the rupturing
  surfaces.
\newblock {\em Journal of Geophysical Research}, 104(B1):817--844.
\newblock \url{https://doi.org/10.1029/1998JB900007}.

\bibitem[Okubo and Dieterich, 1984]{OkuDiet84}
Okubo, P.~G. and Dieterich, J.~H. (1984).
\newblock Effects of physical fault properties on frictional instabilities
  produced on simulated faults.
\newblock {\em Journal of Geophysical Research}, 89(B7):5817--5827.
\newblock \url{https://doi.org/10.1029/JB089iB07p05817}.

\bibitem[Passel{\`e}gue et~al., 2017]{PaLa17}
Passel{\`e}gue, F.~X., Latour, S., Schubnel, A., Nielsen, S., Bhat, H.~S., and
  Madariaga, R. (2017).
\newblock Influence of fault strength on precursory processes during laboratory
  earthquakes.
\newblock In Thomas, M.~Y., Mitchell, T.~M., and Bhat, H.~S., editors, {\em
  Fault Zone Dynamic Processes: Evolution of Fault Properties During Seismic
  Rupture}.
\newblock \url{https://doi.org/10.1002/9781119156895.ch12}.

\bibitem[Ranjith and Rice, 1999]{RaRi99}
Ranjith, K. and Rice, J.~R. (1999).
\newblock Stability of quasi-static slip in a single degree of freedom elastic
  system with rate and state dependent friction.
\newblock {\em Journal of the Mechanics and Physics of Solids}, 47:1207--1218.
\newblock \url{https://doi.org/10.1016/S0022-5096(98)00113-6}.

\bibitem[Rice and Ruina, 1983]{RiRu83}
Rice, J. and Ruina, A.~L. (1983).
\newblock Stability of steady frictional slipping.
\newblock {\em Journal of applied mechanics}, 50(2):343--349.
\newblock \url{https://doi.org/10.1115/1.3167042}.

\bibitem[Rice and Tse, 1986]{RiTse86}
Rice, J.~R. and Tse, S.~T. (1986).
\newblock Dynamic motion of a single degree of freedom system following a rate
  and state dependent friction law.
\newblock {\em Journal of Geophysical Research}, 91(B1):521--530.
\newblock \url{https://doi.org/10.1029/JB091iB01p00521}.

\bibitem[Rubin and Ampuero, 2005]{RuAm05}
Rubin, A.~M. and Ampuero, J.-P. (2005).
\newblock Earthquake nucleation on (aging) rate and state faults.
\newblock {\em Journal of Geophysical Research: Solid Earth}, 110(B11).
\newblock \url{https://doi.org/10.1029/2005JB003686}.

\bibitem[Ruina, 1983]{Ruin83}
Ruina, A. (1983).
\newblock Slip instability and state variable friction laws.
\newblock {\em Journal of Geophysical Research: Solid Earth},
  88(B12):10359--10370.
\newblock \url{https://doi.org/10.1029/JB088iB12p10359}.

\bibitem[Tullis and Weeks, 1986]{TuWe86}
Tullis, T.~E. and Weeks, J.~D. (1986).
\newblock Constitutive behavior and stabilty of frictional sliding of granite.
\newblock {\em Pure Appl. Geophys.}, 124(3):383--414.
\newblock \url{https://doi.org/10.1007/BF00877209}.

\bibitem[Uenishi and Rice, 2003]{UeRi03}
Uenishi, K. and Rice, J.~R. (2003).
\newblock Universal nucleation length for slip-weakening rupture instability
  under nonuniform fault loading.
\newblock {\em Journal of Geophysical Research: Solid Earth}, 108(B1).
\newblock \url{https://doi.org/10.1029/2001JB001681}.

\bibitem[Viesca, 2016a]{Vies16b}
Viesca, R.~C. (2016a).
\newblock Self-similar slip instability on interfaces with rate-and
  state-dependent friction.
\newblock {\em Proc. Roy. Soc. Lond. A}, 472(2192):20160254.
\newblock \url{https://doi.org/10.1098/rspa.2016.0254}.

\bibitem[Viesca, 2016b]{Vies16a}
Viesca, R.~C. (2016b).
\newblock Stable and unstable development of an interfacial sliding
  instability.
\newblock {\em Physical Review E}, 93(6):060202.
\newblock \url{https://link.aps.org/doi/10.1103/PhysRevE.93.060202}.

\bibitem[Viesca, 2023]{Viesca23}
Viesca, R.~C. (2023).
\newblock Frictional state evolution laws and the non-linear nucleation of
  dynamic shear rupture.
\newblock {\em Journal of the Mechanics and Physics of Solids}, 173(105221).
\newblock \url{https://doi.org/10.1016/j.jmps.2023.105221}.

\bibitem[Viesca and Garagash, 2018]{ViGa18}
Viesca, R.~C. and Garagash, D.~I. (2018).
\newblock Numerical methods for coupled fracture problems.
\newblock {\em Journal of the Mechanics and Physics of Solids}, 113:13--34.
\newblock \url{https://doi.org/10.1016/j.jmps.2018.01.008}.

\bibitem[Xu et~al., 2018]{XuFu18}
Xu, S., Fukuyama, E., Yamashita, F., Mizoguchi, K., Takizawa, S., and Kawakata,
  H. (2018).
\newblock Strain rate effect on fault slip and rupture evolution: {I}nsigth
  from meter-scale rock friction experiments.
\newblock {\em Tectonophysics}, 733:209--231.
\newblock \url{https://doi.org/10.1016/j.tecto.2017.11.039}.

\end{thebibliography}
\end{document}